\pdfoutput=1
\documentclass[11pt,twoside,a4paper,cmspaper,final,collab]{cms-tdr}
\def\svnVersion{d4340e2}\def\svnDate{2022/04/28}\def\cmsCernNoTag{CERN-EP-2021-240}\def\cmsCernDate{\today}\def\cmsMessage{Published in Physical Review D as \href{http://dx.doi.org/10.1103/PhysRevD.105.072008}{\doi{10.1103/PhysRevD.105.072008}.}}
\begin{document}\cmsNoteHeader{SMP-18-011}

\newcommand{\cmsTable}[1]{\resizebox{\textwidth}{!}{#1}}

\newcommand{\pp}{\ensuremath{\Pp\Pp}\xspace}
\newcommand{\mW}{\ensuremath{m_{\PW}}\xspace} 
\newcommand{\had}{\ensuremath{\text{h}}\xspace}
\newcommand{\jet}{\ensuremath{\text{j}}\xspace}

\newcommand{\Wtolnu}{\ensuremath{\PW\to\ell\PAGn}\xspace}
\newcommand{\Wtoqq}{\ensuremath{\PW\to\qqbar'}\xspace}
\newcommand{\Wtoenu}{\ensuremath{\PW\to\Pe\PAGne}\xspace}
\newcommand{\Wtomunu}{\ensuremath{\PW\to\PGm\PAGnGm}\xspace}
\newcommand{\Wtotaunu}{\ensuremath{\PW\to\PGt\PAGnGt}\xspace}
\newcommand{\brWtolnu}{\ensuremath{\mathcal{B}(\Wtolnu)}\xspace}
\newcommand{\brWtoqq}{\ensuremath{\mathcal{B}(\Wtoqq)}\xspace}
\newcommand{\brWtoenu}{\ensuremath{\mathcal{B}(\Wtoenu)}\xspace}
\newcommand{\brWtomunu}{\ensuremath{\mathcal{B}(\Wtomunu)}\xspace}
\newcommand{\brWtotaunu}{\ensuremath{\mathcal{B}(\Wtotaunu)}\xspace}

\newcommand{\Rtaul}{\ensuremath{R_{\PGt/\ell}}\xspace}
\newcommand{\Rtaue}{\ensuremath{R_{\PGt/\Pe}}\xspace}
\newcommand{\Rtaumu}{\ensuremath{R_{\PGt/\PGm}}\xspace}

\newcommand{\Vij}{\ensuremath{V_{ij}}\xspace}
\newcommand{\Vcs}{\ensuremath{V_{\PQc\PQs}}\xspace}
\newcommand{\absVij}{\ensuremath{\abs{\Vij}}\xspace}
\newcommand{\alpSmW}{\ensuremath{\alpS(\mW^{2})}\xspace} 

\newcommand{\WW}{\ensuremath{\PW\PW}\xspace}
\newcommand{\WZ}{\ensuremath{\PW\PZ}\xspace}
\newcommand{\ZZ}{\ensuremath{\PZ\PZ}\xspace}
\newcommand{\tW}{\ensuremath{\PQt\PW}\xspace}
\newcommand{\Wjets}{\ensuremath{\PW+\text{jets}}\xspace}
\newcommand{\Zjets}{\ensuremath{\PZ+\text{jets}}\xspace}
\newcommand{\gjets}{\ensuremath{\Pgg+\text{jets}}\xspace}

\newlength\cmsTabSkip\setlength{\cmsTabSkip}{1ex}

\newlength\cmsFigWidth
\ifthenelse{\boolean{cms@external}}{\setlength\cmsFigWidth{0.49\textwidth}}{\setlength\cmsFigWidth{0.80\textwidth}} 
\ifthenelse{\boolean{cms@external}}{\providecommand{\cmsLeft}{upper\xspace}}{\providecommand{\cmsLeft}{left\xspace}}
\ifthenelse{\boolean{cms@external}}{\providecommand{\cmsRight}{lower\xspace}}{\providecommand{\cmsRight}{right\xspace}}

\cmsNoteHeader{SMP-18-011} 
\title{Precision measurement of the \texorpdfstring{\PW}{W} boson decay branching fractions in proton-proton collisions at \texorpdfstring{$\sqrt{s}=13\TeV$}{sqrt(s) = 13 TeV}}

\date{\today}

\abstract{
The leptonic and inclusive hadronic decay branching fractions of the \PW boson are measured using proton-proton collision data collected at $\sqrt{s}=13\TeV$ by the CMS experiment at the CERN LHC, corresponding to an integrated luminosity of 35.9\fbinv. Events characterized by the production of one or two \PW bosons are selected and categorized based on the multiplicity and flavor of reconstructed leptons, the number of jets, and the number of jets identified as originating from the hadronization of \PQb quarks. A binned maximum likelihood estimate of the \PW boson branching fractions is performed simultaneously in each event category. The measured branching fractions of the \PW boson decaying into electron, muon, and tau lepton final states are $(10.83 \pm 0.10)\%$, $(10.94 \pm 0.08)\%$, and $(10.77 \pm 0.21)\%$, respectively, consistent with lepton flavor universality for the weak interaction. The average leptonic and inclusive hadronic decay branching fractions are estimated to be $(10.89 \pm 0.08)\%$ and $(67.32 \pm 0.23)\%$, respectively. Based on the hadronic branching fraction, three standard model quantities are subsequently derived: the sum of squared elements in the first two rows of the Cabibbo--Kobayashi--Maskawa (CKM) matrix $\sum_{ij}\abs{\Vij}^{2} = 1.984 \pm 0.021$, the CKM element $\abs{\Vcs} = 0.967 \pm 0.011$, and the strong coupling constant at the \PW boson mass scale, $\alpSmW = 0.095 \pm 0.033$.
}

\hypersetup{
    pdfauthor={CMS Collaboration},
    pdftitle    = {Precision measurement of the W boson decay branching fractions in proton-proton collisions at sqrt(s) = 13 TeV},
    pdfsubject  = {CMS},
    pdfkeywords = {CMS, standard model}
}

\maketitle

\section{Introduction}
\label{sec:introduction}

Measurements of the leptonic and hadronic widths of the \PW boson, $\Gamma(\Wtolnu)$ with $\ell = \Pe, \PGm, \PGt$ and $\Gamma(\Wtoqq)$, respectively, or their corresponding decay branching fractions derived from their ratio to the total \PW width, $\mathcal{B}(\Wtolnu,\qqbar')\,=\,\Gamma(\Wtolnu,\qqbar')/\Gamma_\text{\PW,total}$, provide a compelling testing ground to investigate fundamental aspects of the standard model (SM). Primarily, all electroweak (EW) bosons are assumed to couple equally to all three lepton generations, a property known as lepton flavor universality (LFU), and experimental evidence of a departure from this assumption would be a sign of new physics.  In recent years, hints of potential LFU violation have been reported, \eg, in semileptonic decays of \PB mesons where the bottom quark converts into a strange quark through an intermediate \PW boson~\cite{BaBar:2012obs,BaBar:2013mob,LHCb:2015gmp,Belle:2016ure,LHCb:2017vlu}. In addition, other hints of LFU failure have been seen in rarer (electroweak, loop-induced) \PB-meson decays~\cite{Aaij:2019wad,LHCb:2021trn}.
A complementary test of LFU can be carried out by comparing the three branching fractions of the \PW boson in the electron, muon, and tau lepton decay channels. The most precise values of the $\mathcal{B}(\PW\to\ell\PAGn_{\ell})$ fractions have been obtained from combinations of measurements performed by each of the four LEP experiments at CERN~\cite{Schael:2013ita,PDG2020}. Based on these results, a ratio between branching fractions has been obtained,
\begin{linenomath} 
    \begin{equation}
    \label{eq:BRs_LEP}
    \ifthenelse{\boolean{cms@external}}
    {
    \begin{aligned}
        R_{\PGt/(\Pe+\PGm)} &= \frac{2\, \brWtotaunu}{\brWtoenu + \brWtomunu} \\ &= 1.066 \pm 0.025 ,
    \end{aligned}
    }{
    R_{\PGt/(\Pe+\PGm)} = \frac{2\, \brWtotaunu}{\brWtoenu + \brWtomunu} = 1.066 \pm 0.025 ,
    }
    \end{equation}
\end{linenomath}
which shows a 2.6 standard-deviations departure from the SM expectation of $\Rtaul = 0.9996$~\cite{Denner:1991kt,Rtau,dEnterria:2020cpv}. Confirmation of this hint of LFU violation requires more precise measurements of the \PW boson branching fractions than available at LEP. In  proton-proton (\pp) collisions at the LHC, the large cross section for the production of top quark-antiquark pairs (\ttbar), each decaying into a \PW boson and a bottom (\PQb) quark, offers a sizable high-purity sample of \PW boson pairs useful for a high-precision study of their decays. A recent measurement by the ATLAS Collaboration took advantage of the large \ttbar production at the LHC to measure the ratio $\Rtaumu$ by fitting the transverse impact parameter distribution of the \PW-decay muons~\cite{Aad:2020ayz}. The resulting value of $\Rtaumu = 0.992 \pm 0.013$ is in tension with the LEP result, and favors the LFU hypothesis. Measurements of the ratio of the electronic to muonic branching fractions of the \PW boson have also been performed by D0~\cite{D0:1995gzy}, CDF~\cite{CDF:2005bdv}, ATLAS~\cite{ATLAS:2016nqi}, and LHCb~\cite{LHCb:2016zpq}, where each experiment observed values consistent with LFU.

A second motivation to study \PW boson decays arises from the fact that within the SM the \PW hadronic width depends on various free parameters of the theory---such as the strong coupling constant at the \PW mass, $\alpSmW$, and the quark flavor mixing elements of the first two rows of the Cabibbo--Kobayashi--Maskawa (CKM) matrix---that can thereby be indirectly determined. Theoretically, the decay width of the \PW boson into (massless) quarks is provided by the expression,
\begin{linenomath} 
    \ifthenelse{\boolean{cms@external}}
    {
    \begin{multline}
        \label{eq:W_had}
        \Gamma(\Wtoqq) = \frac{\sqrt{2} G_\mathrm{F}\,N_\mathrm{c}}{12 \pi} m_\PW^3 \sum_{i,j} \absVij^2\,\\ \times\Bigl[1+ \sum_{{k}=1}^4 c^{(i)}_\mathrm{QCD}\left(\frac{\alpS}{\pi}\right)^{{k}} + \delta_\mathrm{EW} (\alpha) + \delta_\text{mix} (\alpha \alpS)\Bigr],
    \end{multline}
    }{
    \begin{equation}
        \label{eq:W_had}
        \Gamma(\Wtoqq) = \frac{\sqrt{2} G_\mathrm{F}\,N_\mathrm{c}}{12 \pi} m_\PW^3 \sum_{i,j} \absVij^2\,\Bigl[1+ \sum_{{k}=1}^4 c^{(i)}_\mathrm{QCD}\left(\frac{\alpS}{\pi}\right)^{{k}} + \delta_\mathrm{EW} (\alpha) + \delta_\text{mix} (\alpha \alpS)\Bigr],
        \end{equation}
    }
\end{linenomath}
where the factor before the parentheses is the Born width, which depends on the number of colors $N_\mathrm{c}=3$, the Fermi constant $G_\mathrm{F}$, \mW, and the sum of squared CKM matrix elements \Vij (excluding terms involving the top quark that are not kinematically accessible). The terms in parenthesis of Eq.~(\ref{eq:W_had}) include the higher-order perturbative quantum chromodynamics (QCD) corrections, given by an expansion in $\alpS^{{k}}$ coefficients known up to order ${k}=4$~\cite{Baikov:2008jh}, the EW corrections $\delta_{\mathrm{EW}}$ known to order $\mathcal{O}(\alpha)$~\cite{Denner:1991kt} (where $\alpha$ is the electromagnetic coupling), and the mixed EW plus QCD corrections $\delta_{\text{mix}}$ known to order $\mathcal{O}(\alpha \alpS)$~\cite{Kara:2013dua}. Based on Eq.~(\ref{eq:W_had}) and the ratio of hadronic to leptonic branching fractions of the \PW boson, the unitarity of the first two rows of the CKM matrix can be tested by searching for a deviation from $\sum_{\PQu,\PQc,\PQd,\PQs,\PQb} \abs{\Vij}^2 \equiv 2$. Additionally, it is possible to indirectly determine the value of $\abs{\Vcs}$~\cite{Schael:2013ita,dEnterria:2016rbf}, which currently has the largest absolute uncertainty among the elements of the first two rows of the CKM matrix. Based on the current world-average values of the CKM elements~\cite{PDG2020}, the quadratic sum of the elements in the first two CKM matrix rows can be derived, $\sum_{\PQu,\PQc,\PQd,\PQs,\PQb} \abs{V_{{ij}}}^2 = 2.002~\pm~0.027$, with a 1.3\% precision dominated by the uncertainty of the $\abs{\Vcs}$ element. Consequently, a measurement of the inclusive \PW hadronic branching fraction with subpercent uncertainties provides a more precise, albeit indirect, determination of the value of the $\abs{\Vij}^{2}$ sum as well as of $\abs{\Vcs}$. Assuming CKM unitarity, it is also possible to determine the value of \alpSmW via Eq.~(\ref{eq:W_had}), although not with a precision competitive with other extractions to date~\cite{PDG2020,dEnterria:2020cpv}. 

This paper describes a measurement of the three leptonic branching fractions, as well as of the inclusive hadronic branching fraction, of \PW boson decays. The analysis is based on \pp collision data at a center-of-mass energy of 13\TeV  corresponding to an integrated luminosity of 35.9\fbinv~\cite{CMS-LUM-17-003} collected by the CMS experiment at the CERN LHC in 2016. Selected events are required to contain at least one electron or muon with large transverse momentum, \pt. The events are grouped into final-state categories that primarily target decays of two \PW bosons originating from \ttbar production. The values of the \PW boson branching fractions are estimated from a binned maximum likelihood fit to data in final states selected based on the number and the flavor of leptons, the number of jets, the number of those jets identified as originating from \PQb quarks, and a category-dependent kinematic variable.

\section{The CMS detector}
\label{sec:cms}

The central feature of the CMS apparatus is a superconducting solenoid, 13\unit{m} in length and 6\unit{m} in diameter, which provides an axial magnetic field of 3.8\unit{T}. Within the field volume there are several particle detection systems. Charged particle trajectories are measured by silicon pixel and strip trackers, covering $0 < \phi < 2\pi$ in azimuth and $\abs{\eta} < 2.5$ in pseudorapidity,  where $\eta$ is defined as $-\log[\tan(\theta/2)]$ and $\theta$ is the polar angle of the trajectory of the particle with respect to the counterclockwise proton beam direction. A lead tungstate crystal electromagnetic calorimeter (ECAL) and a brass and scintillator hadron calorimeter (HCAL) surround the tracking volume and cover the region $\abs{\eta} < 3$. The calorimeters provide energy measurements of photons, electrons, and jets of hadrons. A lead and silicon strip preshower detector is located in front of the ECAL endcap. Muons are identified and measured in gas-ionization detectors embedded in the steel flux return yoke outside of the solenoid. The detector is nearly hermetic, allowing energy balance measurements in the plane transverse to the beam direction. A more detailed description of the CMS detector is reported in Ref.~\cite{Chatrchyan:2008zzk}. 

\section{Simulated event samples}
\label{sec:datasets}

Simulated Monte Carlo (MC) event samples are generated for the processes defined as signal (\ttbar, \tW, \WW, and \Wjets) and backgrounds (\Zjets, \gjets, $\PW\PZ$, and $\PZ\PZ$). The contribution to the background originating from QCD multijet production is estimated using control samples in data. The \POWHEG v2~\cite{Alioli:2010xd, Kardos:2014dua, Frixione:2002ik, Nason:2004rx, Re:2010bp} MC event generator is used at next-to-leading order (NLO) QCD accuracy to produce samples of $\ttbar$, single top quark produced in association with a \PW boson (\tW), and most of the relevant diboson processes (\WW, \WZ, and $\ZZ\to 2\ell 2\PGn$). The \Wjets MC samples are generated at leading-order (LO) QCD accuracy using the \MADGRAPH event generator~\cite{MadGraph}. Drell--Yan, \gjets, \WZ, and semileptonic \ZZ decay modes are generated at NLO QCD accuracy with \MGvATNLO~\cite{MadGraphNLO,Frixione:2007vw}. In all cases, the MC samples are obtained with the NNPDF3.0 parton distribution functions (PDFs), and are interfaced with \PYTHIA 8.212~\cite{Sjostrand:2014zea, Skands:2014pea} for parton showering and hadronization. The underlying event (UE) \PYTHIA 8.212 tune used for most samples is CUETP8M1~\cite{Khachatryan:2015pea} with the exception of the \ttbar case which uses the dedicated CUETP8M2T4 tune~\cite{CMS-PAS-TOP-16-021}. The CMS detector response is simulated with a \GEANTfour-based model~\cite{Agostinelli:2002hh}, and the events are reconstructed and analyzed using the same software employed to process collision data. 

The impact of pileup \pp collisions on the event reconstruction~\cite{CMS:2020ebo} is accounted for in simulation by superimposing simulated minimum bias \pp events on top of each process of interest. Because the distribution of the number of pileup events in the original simulation is not the same as in data, the former is reweighted to match the latter. Scale factors are also applied to account for differences between data and simulation with respect to modeling of the trigger efficiencies, as well as lepton reconstruction, identification, and isolation efficiencies. Additional corrections are applied to account for the energy scale and \pt resolution of charged leptons. The jet energy scale (JES), resolution (JER), and \PQb tagging efficiency and multivariate discriminator distributions measured in data are used to correct the simulated events.

LFU is assumed by default in the simulated event samples, taking $\brWtolnu = 10.86\%$ for each leptonic decay mode~\cite{PDG2020}. For the \PGt decays, its hadronic and leptonic branching fractions are taken from their current world-average values~\cite{PDG2020}. 

\section{Event selection and reconstruction}
\label{sec:selection}

A two-tier trigger system~\cite{Sirunyan:2020zal,Khachatryan:2016bia} selects \pp collision events of interest for physics analysis. The triggers used to collect data require the detection of a single muon (electron) with $\pt > 24\,(27)\GeV$ and $\abs{\eta} < 2.4\,(2.5)$.

Though the selection is designed mainly to collect events originating from \ttbar production, the chosen criteria also accept contributions from \tW, \WW and \Wjets production, which are thereby also considered as signal processes in this analysis. The background processes include the production of multiple QCD jets, \PZ boson plus jets, and \WZ and \ZZ dibosons. The \WZ production is not considered as part of the signal processes because of its negligible contribution. The selection of events consistent with the signal processes requires reconstructing electrons, muons, hadronically decaying \PGt leptons (\tauh), and hadronic jets. Additionally, to suppress backgrounds it is useful to determine whether reconstructed jets originate from the fragmentation of \PQb quarks. 

A global particle-flow (PF) event reconstruction~\cite{Sirunyan:2017ulk} is used to reconstruct and identify each individual particle in a \pp collision, with an optimized combination of all subdetector information. Photons are identified as ECAL energy clusters not linked to the extrapolation from any charged particle trajectory reconstructed in the tracker. Electrons are identified as a primary charged particle track plus, potentially, any ECAL energy clusters matched to the track as well as to any bremsstrahlung photons emitted along the way through the tracker material. Muons are identified as tracks in the central tracker that are consistent with either a track or several hits in the muon system, and associated with calorimeter deposits compatible with the muon hypothesis. Charged hadrons are identified as charged particle tracks neither identified as electrons, nor as muons. Finally, neutral hadrons are identified as HCAL energy clusters not linked to any charged hadron trajectory, or as ECAL and HCAL signals with energies above those expected to be deposited by a charged hadron.

The candidate vertex with the largest value of summed physics-object $\pt^2$ is the primary \pp interaction vertex (PV). The physics objects are the jets, clustered using the anti-\kt jet finding algorithm~\cite{Cacciari:2008gp,Cacciari:2011ma} with the tracks assigned to candidate vertices as inputs, and the associated missing transverse momentum, \ptmiss, taken as the negative vector \pt sum of all jets. Quality requirements are applied to reconstructed PVs to guarantee that they come from a hard scattering event~\cite{Chatrchyan:2014fea}.

Electrons are reconstructed by combining information from the ECAL and the tracking system using a Gaussian-sum filter method~\cite{Baffioni:2006cd}. Electrons are required to have $\pt > 10\GeV$, and lie within the geometrical acceptance of $\abs{\eta} < 2.5$. Corrections are applied to account for mismeasurements of the electron momentum scale and resolution. To select electrons that have originated from the prompt decay of an EW boson, an isolation variable is constructed by summing the \pt of charged hadrons ($I_\mathrm{ch}$), neutral hadrons ($I_\mathrm{neu}$), and photons ($I_{\gamma}$) within a cone of radius $\Delta R = \sqrt{\smash{(\Delta\eta)^2}+\smash{(\Delta\phi)}^2} = 0.4$ around the electron candidate direction, and subtracting the contribution from pileup. The combined PF isolation for electron candidates is defined as,
\begin{linenomath}
    \begin{equation} 
        I_\mathrm{PF} = I_\text{ch} + \max\left(0, I_\text{neu} + I_{\gamma} - \rho A_\text{eff}(\eta_{\Pe}) \right),
    \end{equation}
\end{linenomath}
where the pileup correction $\rho A_\text{eff}$ depends on the the median transverse energy density per unit area in the event $\rho$, and on the area of the isolation region $A_\text{eff}(\eta_{\Pe})$ weighted by a factor that accounts for the $\eta$ dependence of the pileup transverse energy density around the electron~\cite{Khachatryan:2015hwa}. Electrons reconstructed in the barrel ($\abs{\eta} < 1.479$) or endcap ($\abs{\eta} > 1.479$), are required to have $I_\mathrm{PF}/\pt^\Pe < 0.0588$ and $0.0571$, respectively. 

Muon candidates are reconstructed using both the muon and tracker detector subsystems. The coverage of these two detector systems allows reconstruction of muons within $\abs{\eta} < 2.4$ and with \pt as low as 5\GeV~\cite{Chatrchyan:2012xi}. Muons are required to be reconstructed by both the global and tracker reconstruction algorithms. These algorithms are distinct in that the tracker \PGmpm reconstruction begins with tracker information and extrapolates the trajectory to find consistency with hits in the muon system, whereas the global muon algorithm inverts the reconstruction steps starting from the muon system and finding trajectories in the tracker that are consistent with them. The combination of these two algorithms results in a muon reconstruction that is efficient in detecting muons within the detector acceptance as well as accurate in predicting their momenta.

For the purpose of selecting muons promptly produced from weak boson decays, additional identification and isolation requirements are applied~\cite{Sirunyan:2018fpa}. The muon identification requirements are designed to have a high selection efficiency and a low probability of misidentification against nonprompt muons. The isolation of muons is defined as the scalar \pt sum of all charged-hadron, neutral-hadron, and photon PF candidates in a cone of radius $\Delta R = 0.4$ around the $\PGm$ direction. Th isolation includes a term ($I_\mathrm{pileup}$) accounting for neutral particles produced by overlapping \pp collisions by subtracting half the average energy deposited by pileup,
\begin{linenomath}
    \begin{equation} 
        I_\mathrm{PF} = I_\text{ch} + \max\left(0, I_\text{neu} + I_{\gamma} - 0.5 I_\text{pileup}\right).
    \end{equation}
\end{linenomath}
All muons are required to have $I_\mathrm{PF}/\pt^\PGm < 0.15$, except when an isolation sideband is used to estimate backgrounds.

Hadronically decaying \PGt leptons ($\tauh$) are reconstructed using the hadron-plus-strips algorithm~\cite{Sirunyan:2018pgf}. This algorithm reconstructs \tauh candidates seeded by a PF jet that is consistent with either a single or a triple charged-pion decay of the \PGt lepton. In the single charged-pion decay mode, additional neutral pions are reconstructed using their diphoton decays. Any \tauh that overlaps with reconstructed muons or electrons is rejected. Jets not originating from \PGt lepton decays are rejected by a multivariate discriminator that takes into account the pileup contribution to the neutral component of the \PGt lepton decay~\cite{Sirunyan:2018pgf}. The reconstructed \tauh are required to have $\pt > 20\GeV$ and $\abs{\eta} < 2.3$. A working point with an identification efficiency of ${\approx}50\%$ and a misidentification efficiency of ${\approx}0.2\%$ is used in selecting \tauh candidates. 
Scale factors are derived to account for differences between \tauh identification efficiencies in simulation compared with data~\cite{Sirunyan:2018pgf} in two control regions enriched in \PZ and \ttbar production. The differences of the reconstructed \tauh energy between data and simulation are also corrected in simulation using scale factors determined in a $\PZ\to\PGt\PGt$ region.

Jets are reconstructed from PF candidates~\cite{Sirunyan:2017ulk} clustered using the anti-\kt algorithm with a distance parameter of $0.4$, and are required to have $\pt > 30\GeV$ and $\abs{\eta} < 2.4$. Jets are corrected to account for pileup contamination, differences in absolute response of jet \pt between data and simulation, and relative response in $\eta$~\cite{ref:jetscale}. To reduce contamination from photons and prompt leptons, additional identification requirements are applied to the jets. Jets are vetoed if they overlap,  within a cone of radius $\Delta R = 0.4$ around the jet direction, with any reconstructed muon, electron, or \tauh lepton passing the identification requirements described above.

Jets originating from the hadronization of \PQb quarks are identified using the combined secondary vertex \PQb tagging algorithm~\cite{Sirunyan:2017ezt} that uses secondary, displaced vertices and track lifetime information. The \PQb-tagged jets are selected such that their detection efficiency is 63\% for a 1\% misidentification rate. To account for the difference in \PQb tagging efficiency between data and simulation, \pt-dependent scale factors are used to modify the \PQb-tag status of individual jets in simulation depending on whether the jet originates from a \PQb quark, a \PQc quark, or a light quark or gluon.

\section{Event categorization}
\label{sec:evt_classif}

Requirements are applied offline to categorize events based on the multiplicity of reconstructed leptons, jets, and \PQb-tagged jets passing a minimum \pt threshold, as summarized in Table~\ref{tab:event_categories}. In categories with two leptons in the final state, the leptons are required to have opposite-sign electric charges. Events in the $\Pe\Pe$ and $\PGm\PGm$ categories are rejected if the lepton pair invariant mass is between 75 and 105\GeV in order to reduce the contamination from \PZ boson events. The various categories are dominated by \PW decays originating from \ttbar production (90\%) with minor contributions from \tW (4.4\%), \WW (1.4\%), and \Wjets (4.2\%) processes, whereas the background consists mainly of Drell--Yan and multijet QCD production, with almost negligible contributions from \WZ and \ZZ diboson processes. 

\begin{table*}
    \topcaption{Categorization of events based on the triggering lepton, the number of reconstructed and selected leptons ($N_{\Pe}$, $N_{\PGm}$, $N_{\tauh}$), number of jets ($N_\jet$), and number of \PQb-tagged jets ($N_{\PQb}$). Kinematic requirements of the leptons and jets are listed in the fourth column. Categories with two leptons in the final state require the selected leptons to have opposite signs. The second-to-last column lists the targeted \PW boson branching fractions, and the last column provides the approximate number of \PW decays collected in each category. 
    \label{tab:event_categories}}
    \centering                                                              
    \renewcommand{\arraystretch}{1.5}                                       
    \resizebox{\textwidth}{!}{                                           
    \begin{scotch}{c c ccccc l l c}                                            
        Trigger               & Label             & $N_{\Pe}$ & $N_{\PGm}$ & $N_{\tauh}$ & $N_{\jet}$ & $N_{\PQb}$ & Kinematic requirements          & Target \PW boson & Approx.\ num.       \\
        & & & & & & & & branching fractions & of \PW decays \\
        \hline                                                                                      
        \multirow{4}{*}{\Pe}  & $\Pe\Pe$          & 2         & 0          & 0           & $\geq$2       & $\geq$1    & $\pt^{\Pe} > 30, 20\GeV$, $\abs{m_{\Pe\Pe} - m_{\PZ}} > 15\GeV$  & $\PW\to\Pe\PAGne,\PGt\PAGnGt$ & $1.1 \times 10^{5}$ \\
                              & $\Pe\PGm$         & 1         & 1          & 0           & $\geq$0       & $\geq$0    & $\pt^{\Pe} > 30\GeV$, $\pt^{\PGm} > 10\GeV$  &  $\PW\to\Pe\PAGne,\PGm\PAGnGm,\PGt\PAGnGt$ & $4 \times 10^{5}$ \\
                              & $\Pe\tauh$        & 1         & 0          & 1           & $\geq$0       & $\geq$0    & $\pt^{\Pe} > 30\GeV$, $\pt^{\tauh} > 20\GeV$   &  $\PW\to\Pe\PAGne,\PGt\PAGnGt$ & $8 \times 10^{4}$ \\
                              & $\Pe \had$  & 1         & 0          & 0           & $\geq$4       & $\geq$1    & $\pt^{\Pe} > 30\GeV$, $\pt^{\jet} > 30\GeV$ &  $\PW\to\Pe\PAGne,\qqbar'$  & $1.4 \times 10^{6}$ \\[1.5\cmsTabSkip]

        \multirow{4}{*}{\PGm} & $\PGm\Pe$         & 1         & 1          & 0           & $\geq$0       & $\geq$0    & $\pt^{\PGm} > 25\GeV$, $\pt^{\Pe} > 20\GeV$        & $\PW\to\Pe\PAGne,\PGm\PAGnGm,\PGt\PAGnGt$  & $2 \times 10^{5}$   \\
                              & $\PGm\PGm$        & 0         & 2          & 0           & $\geq$2       & $\geq$1    & $\pt^{\PGm} > 25, 10\GeV$, $\abs{m_{\PGm\PGm} - m_{\PZ}} > 15\GeV$ & $\PW\to\PGm\PAGnGm,\PGt\PAGnGt$ & $3 \times 10^{5}$   \\
                              & $\PGm\tauh$       & 0         & 1          & 1           & $\geq$0       & $\geq$0    & $\pt^{\PGm} > 25\GeV$, $\pt^{\tauh} > 20\GeV$       & $\PW\to\PGm\PAGnGm,\PGt\PAGnGt$            & $1.3 \times 10^{5}$ \\
                              & $\PGm \had$ & 0         & 1          & 0           & $\geq$4       & $\geq$1    & $\pt^{\PGm} > 25\GeV$, $\pt^{\jet} > 30\GeV$     & $\PW\to\PGm\PAGnGm,\qqbar'$ & $2.1 \times 10^{6}$ \\
    \end{scotch}}                                                          
\end{table*}

Each of the categories is designed to target particular combinations of \PW decay modes, but will include events attributable to different decays. The selection categories mostly contain events collected using only one of the triggers with the exception of the $\Pe\PGm$ and $\PGm\Pe$ categories where overlap is accounted for by rejecting any duplicated events. Because \PGt leptons are not detected directly, but through their decay products, all categories contain a mixture of events with final states that include electrons, muons, or jets originating either directly from \PW boson decays or through intermediate \PGt decays. This ambiguity in reconstruction is maximal in the $\Pe\tauh$ and $\PGm\tauh$ categories with two or more jets, because of the higher probability of a jet originating from a \PW boson decay being misidentified as originating from a \PGt lepton decay. The categories denoted by $\Pe\had$ and $\PGm\had$ are intended to target decay modes where one of the \PW bosons has decayed to quarks.

To further improve the sensitivity to specific branching fractions and constrain some of the systematic uncertainties, events are further categorized based on the jet and \PQb tag multiplicities as shown in Table~\ref{tab:jet_categories}. Events with $N_{\jet} \geq$1 and $N_{\PQb} \geq$1 comprise the bulk of the signal with most of the events originating from \ttbar production. These events also contain some contribution from $\PQt\PW$ production and, in the case of the $\Pe\tauh$, $\PGm\tauh$, $\Pe\had$, and $\PGm\had$ categories, $\PW+$\,jets production. Events in the $\Pe\tauh$ and $\PGm\tauh$ categories with at least one jet that is not \PQb-tagged are used in control regions for the $\tauh$ identification, and include additional requirements to enhance the presence of Drell--Yan events: $40 < m_{\ell\tauh} < 100\GeV$, $\Delta\phi(\ell, \tauh) > 2.5$, and $\mT^\ell < 60\GeV$, where $\mT^\ell$ is the transverse mass of the electron or muon defined as $\mT^\ell = \sqrt{\smash[b]{2 \pt^\ell\ptmiss [1-\cos\Delta\phi(\pt^\ell, \ptmiss)]}}$, where $\Delta\phi(\pt^\ell, \ptmiss)$ is the angle between the electron or muon \pt and the \ptmiss. In general, events with lower jet and \PQb tag multiplicities have larger contributions from background processes and are mainly useful in constraining systematic uncertainties associated with those processes. The exception is in categories with low jet multiplicity and no \PQb tags in $\Pe\PGm$ final states where there is significant contribution from $\PW\PW$ production in addition to background processes. Categories with $\Pe\tauh$ and $\PGm\tauh$ and at least one \PQb tag are also further subdivided depending on whether there are exactly two jets or more than two jets in the event. The reasoning for this choice is based on the fact that events with exactly two jets are more likely to come from events where one \PW boson has decayed to a \PGt lepton and the two jets originated from the \PQb quarks resulting from the top quark decays, whereas events with a third jet are likely to have arisen from a hadronic \PW decay where one jet has been incorrectly reconstructed as a \tauh.

\begin{table}[htpb!]
    \centering
    \setlength{\tabcolsep}{0.4em}
    \renewcommand{\arraystretch}{1.5}
    \topcaption{Categorization of events with electrons, muons, and \tauh passing the reconstruction criteria, based on their jet and \PQb-tagged jet multiplicities, used to define signal-enriched and control regions. Events in the $\Pe\tauh$ and $\PGm\tauh$ categories with at least one jet that is not \PQb-tagged are additionally required to satisfy $40 \leq m_{\ell\tauh} \leq 100 \GeV$, $\Delta\phi(\ell, \tauh) > 2.5$, and $\mT^\ell < 60 \GeV$.
	}
    \label{tab:jet_categories}
    \begin{scotch}{l|c|c|c|c|c}
                                     & $N_{\jet} = 0$       & $N_{\jet} = 1$                               & $N_{\jet} = 2$                               & $N_{\jet} = 3$                           & $N_{\jet} \geq$4 \\
	\hline
    \multirow{2}{*}{$N_{\PQb} = 0$}    & $\Pe\tauh$, $\PGm\tauh$, & $\Pe\tauh$, $\PGm\tauh$,                         & \multicolumn{2}{c}{$\Pe\tauh$, $\PGm\tauh$}      &                                             \\
                                     & $\Pe\PGm$                & $\Pe\PGm$                                        & \multicolumn{2}{c}{$\Pe\PGm$}  &    \\
	\hline
    \multirow{3}{*}{$N_{\PQb} = 1$}    &                          & $\Pe\tauh, \PGm\tauh$, & $\Pe\tauh$, $\PGm\tauh$                          & \multicolumn{2}{c}{$\Pe\tauh$, $\PGm\tauh$} \\
	\cline{4-6}
                                     &  & $\Pe\PGm$ & \multicolumn{3}{c}{$\Pe\Pe, \PGm\PGm, \Pe\PGm$} \\
	\cline{3-6}
                                     & \multicolumn{4}{c|}{}    & $\Pe \had$, $\PGm \had$             \\
	\hline

    \multirow{3}{*}{$N_{\PQb} \geq$2} & \multicolumn{2}{c|}{}    & $\Pe\tauh$, $\PGm\tauh$                          & \multicolumn{2}{c}{$\Pe\tauh$, $\PGm\tauh$}     \\
	\cline{4-6}
                                     & \multicolumn{2}{c|}{}    & \multicolumn{3}{c}{$\Pe\Pe, \PGm\PGm, \Pe\PGm$} \\
	\cline{4-6}
                                     & \multicolumn{4}{c|}{}    & $\Pe \had$, $\PGm \had$             \\
    \end{scotch}
\end{table}

In several of the analysis categories, there is a nonnegligible contamination of nonprompt leptons originating from QCD multijet production. This contamination mainly affects the $\Pe\had$ and $\PGm\had$ decay channels, as well as decays with \tauh candidates in the final state. Two different methods are used for estimating nonprompt-lepton contamination directly from data as explained next. 

To estimate the nonprompt-lepton background originating from multijets in the $\Pe \had$ and $\PGm \had$ categories, a multijet-dominated control region is selected by inverting the lepton isolation requirement. To map the anti-isolated control region into the signal region, transfer factors are determined in a second, orthogonal, control region enriched in $\PW+$\,jets or $\PZ+$\,jets production. These events are tagged by the leptonic decay products of the \PW or \PZ boson, and the additional jets are used to extract the transfer factors from the ratio of the number of leptons passing the nominal isolation requirements to the number passing a looser criterion but failing the nominal, tighter criterion. The transfer factors are determined as a function of the \pt and $\eta$ of the nonprompt lepton, and simulation is used to account for the contamination from processes that produce prompt leptons. The transfer factors are applied as weights to events with the same selection as the signal region but where the leptons pass a loose isolation requirement and fail the tighter requirement used to select signal events.

For event categories with a \tauh candidate, the multijet contribution is estimated from control regions selected by inverting the requirement that the leptons have opposite-sign electric charge. This method relies on the fact that there are few SM processes that give rise to same-sign lepton pair final states, and the events instead originate primarily from misidentification of a hadronic jet or nonprompt lepton as being a prompt lepton. Events gathered in the same-sign control region are scaled by a set of transfer factors determined separately in another, orthogonal, multijet-enriched control region selected by inverting the isolation requirements of the triggering electron or muon. Simulated processes are used to account for contamination from prompt lepton production in all control regions, and mainly include $\PZ\to\PGt\PGt$ (where the \tauh charge is mismeasured) and $\PW+$\,jets. The method is validated in a control region that is enriched in multijets, $\PW+$\,jets, and $\PZ\to\PGt\PGt$ processes selected by requiring no hadronic jets. 

\section{Extraction of branching fractions}
\label{sec:method}

The determination of the \PW branching fractions is carried out using a maximum likelihood estimation (MLE) approach that fits histogram templates, derived from the signal and background estimates, to the data. To explain the method, it is useful to encode the branching fractions into a vector, 
\begin{linenomath}
\begin{equation}
    \beta = \{\beta_{\Pe}, \beta_{\PGm}, \beta_{\PGt}, \beta_\had\},
\end{equation}
\end{linenomath}
where the subscript indicates the decay mode of the \PW boson (all hadronic decay modes, $\had$, are grouped together). Further taking into account the fraction of $\PGt$ decay modes, $\mathbf{t} = \{t_{\Pe}, t_{\PGm}, t_\had\}$, the branching fraction vector can be rewritten,
\begin{linenomath}
\begin{equation}
    \label{eq:br_vector}
    \beta' = \{\beta_{\Pe}, \beta_{\PGm}, \beta_{\PGt}t_{\Pe},
    \beta_{\PGt}t_{\PGm}, \beta_{\PGt}t_\had, \beta_\had\}.
\end{equation}
\end{linenomath}
This parameterization is sufficient for single $\PW$ processes, but because final states with two \PW bosons are of primary interest, it is necessary to consider all possible $\PW$ pair decay combinations. This can be represented by the outer product of $\boldsymbol{\beta'}$ with itself,
\begin{linenomath}
\begin{equation}
\label{eq:br_matrix}
    \mathbf{B} =  \beta' \otimes \beta',
\end{equation}
 \end{linenomath}
that is a 36-element symmetric matrix with 21 unique elements. 

The signal samples mainly consist of events resulting from the decay of two \PW bosons, which are split into 21 categories based on inspecting generator-level event information. The selection and identification efficiencies for the signal samples can be written in a matrix form, with elements corresponding to those in Eq.~(\ref{eq:br_matrix}),
\begin{linenomath}
\begin{equation}
\label{eq:eff_matrix}
\mathbf{E} = \begin{bmatrix}
                \epsilon_{\Pe\Pe}         & \epsilon_{\Pe\PGm}         & \epsilon_{\Pe\PGt_{\Pe}}         & \epsilon_{\Pe\PGt_{\PGm}}         & \epsilon_{\Pe\tauh}         & \epsilon_{\Pe \had}   \\
                \epsilon_{\Pe\PGm}        & \epsilon_{\PGm\PGm}        & \epsilon_{\PGm\PGt_{\Pe}}        & \epsilon_{\PGm\PGt_{\PGm}}        & \epsilon_{\PGm\tauh}        & \epsilon_{\PGm \had}  \\
                \epsilon_{\Pe\PGt_{\Pe}}  & \epsilon_{\PGm\PGt_{\Pe}}  & \epsilon_{\PGt_{\Pe}\PGt_{\Pe}}  & \epsilon_{\PGt_{\Pe}\PGt_{\PGm}}  & \epsilon_{\PGt_{\Pe}\tauh}  & \epsilon_{\PGt_{\Pe}\had}      \\
                \epsilon_{\Pe\PGt_{\PGm}} & \epsilon_{\PGm\PGt_{\PGm}} & \epsilon_{\PGt_{\Pe}\PGt_{\PGm}} & \epsilon_{\PGt_{\PGm}\PGt_{\PGm}} & \epsilon_{\PGt_{\PGm}\tauh} & \epsilon_{\PGt_{\PGm}\had}     \\
                \epsilon_{\Pe\tauh}       & \epsilon_{\PGm\tauh}       & \epsilon_{\PGt_{\Pe}\tauh}       & \epsilon_{\PGt_{\PGm}\tauh}       & \epsilon_{\tauh\tauh}       & \epsilon_{\tauh \had} \\
                \epsilon_{\Pe \had} & \epsilon_{\PGm \had} & \epsilon_{\PGt_{\Pe}\had}           & \epsilon_{\PGt_{\PGm}\had}           & \epsilon_{\tauh \had} & \epsilon_\mathrm{hh}
             \end{bmatrix},
\end{equation}
\end{linenomath}
where the subscript on the $\PGt$ indicates it decays to an electron, a muon, or hadrons. This matrix is constructed for each of the categories described in Tables~\ref{tab:event_categories} and~\ref{tab:jet_categories}, and it is further parameterized as a function of category-dependent observables, such as the subleading lepton \pt. Each individual efficiency in Eq.~(\ref{eq:eff_matrix}) is given by the ratio,
\begin{linenomath}
\begin{equation}
\label{eq:model_eff}
    \epsilon = \frac{\sum_{i}w_{i}}{N_\text{gen}},
\end{equation}
\end{linenomath}
where $w_{i}$ is a weight for each selected event including all scale-factor corrections discussed in Section~\ref{sec:selection}, and $N_\text{gen}$ is the total number of events generated for the process under consideration including generator-level and scale-factor corrections.

The estimated number of events for a given final state (corresponding to the binned kinematic observable `i' and category `j', see below) is then given by,
\begin{linenomath}
\begin{equation}
\label{eq:data_model}
N_{ij} = \sum_{{{k\in \text{sig}}}} \sigma_{k} \Lumi E_{ij}^{k} \mathcal{B}_{ij} + \sum_{l\in {\text{bkg}}} N_l,
\end{equation}
\end{linenomath}
where $\sigma_{k}$ is the cross section of each signal process k that contributes to a given \PW boson decay with branching fraction $\mathcal{B}_{ij}$ and efficiency $E_{ij}^{k}$, $\Lumi$ is the integrated luminosity, and $N_l$ is the predicted number of events for the background process $l$. For \Wjets events, the vector defined in Eq.~(\ref{eq:br_vector}) is used with the corresponding vector of efficiencies for each decay mode. In practical terms, the actual encoded parameterization of Eq.~(\ref{eq:eff_matrix}) includes a free parameter representing the ratio of the branching fraction to the nominal branching fractions used in simulation multiplied by the yield determined from the simulation with the nominal values.

For each category, events are further binned based on a single kinematic observable in each category. The observable is selected to enhance the discrimination between decay products that come directly from the \PW boson decay from those with an intermediate \PGt lepton decay. The variables that are selected for each lepton flavor category are as follows:
\begin{itemize}
    \item $\Pe\Pe$: the subleading electron \pt,
    \item $\PGm\PGm$: the subleading muon \pt,
    \item $\Pe\PGm$: the subleading lepton \pt,
    \item $\Pe\tauh$ and $\PGm\tauh$: the hadronic tau \pt,
    \item $\Pe h$ and $\PGm h$: the lepton \pt.
\end{itemize}
The largest benefit of including this kinematic information comes in the $\Pe\Pe$, $\Pe\PGm$, and $\PGm\PGm$ categories where the light leptons originating from the decay of a $\PGt$ lepton tend to have lower momenta than those originating directly from a \PW boson. 

Templates are generated by binning the data of each category into histograms using the Bayesian block algorithm~\cite{Pollack:2017srh}. The binning is calculated independently for each category based on $10^{4}$ simulated \ttbar events. Effectively, this procedure parameterizes the efficiency matrix in Eq.~(\ref{eq:eff_matrix}) as a function of the extra variables listed above. The predicted yield in each \pt bin ${i}$ and category $j$ is a linear combination of the signal, $s$, and background, $b$, templates given by
\begin{linenomath}
\begin{equation}
    f_{ij}(\boldsymbol{\beta}, \boldsymbol{\theta}) = \sum_{k\in\text{sig}}s_{ij,k}(\boldsymbol{\beta}, \boldsymbol{\theta}) + \sum_{l\in \text{bkg}} b_{ij,l}(\boldsymbol{\theta}),
    \label{eq:templates} 
\end{equation}
\end{linenomath}
where the effects of systematic uncertainties are accounted for by incorporating nuisance parameters (NPs) $\boldsymbol{\theta}$ into the model~\cite{Conway:2011in}, as described in Section~\ref{sec:systematics}. Having constructed the model for the data, the negative log likelihood can then be formulated and minimized for values of the \PW boson branching fractions. Including terms for the NPs, and their prior uncertainty, $\pi(\theta)$, the negative log likelihood is expressed as,
\begin{linenomath}
\begin{equation}
\label{eq:nll_full}
    \ifthenelse{\boolean{cms@external}}
    {
    \begin{aligned}
        L(\boldsymbol{\beta}, \boldsymbol{\theta}) = &\sum_{j\in\text{category}} \sum_{i\in \pt \text{bins}} \bigl[-y_{ij}\ln f_{ij}(\boldsymbol{\beta}, \boldsymbol{\theta}) \\
    &+ f_{ij}(\boldsymbol{\beta}, \boldsymbol{\theta})\bigr] + \sum_{\theta \in \boldsymbol{\theta}}\pi(\theta),
    \end{aligned}
} {
    L(\boldsymbol{\beta}, \boldsymbol{\theta}) = \sum_{j\in\text{category}} \sum_{i\in \pt \text{bins}} \bigl[-y_{ij}\ln f_{ij}(\boldsymbol{\beta}, \boldsymbol{\theta}) + f_{ij}(\boldsymbol{\beta}, \boldsymbol{\theta})\bigr] + \sum_{\theta \in \boldsymbol{\theta}}\pi(\theta),
}
\end{equation}
\end{linenomath}
where $y_{ij}$ is the measured data yield in \pt bin i of category $j$, and $f_{ij}$ are the templates defined in Eq.~(\ref{eq:templates}). The NPs are treated either as affecting the overall normalization of a process in a given channel, or affecting some mixture of the shape of the kinematic distribution being fit and its normalization. For the latter case, morphing templates are generated with the NPs shifted up and down by one standard deviation. The constraints on NPs are assumed to be Gaussian. To reduce the impact of some of the more consequential NPs (\eg, the \tauh candidate reconstruction efficiency), additional control regions in the $\Pe\tauh$ and $\PGm\tauh$ categories enriched in $\PZ\to\PGt\PGt$ events are included in the fit. 

\begin{figure*}[htbp!]
    \centering
    \includegraphics[width=0.49\textwidth]{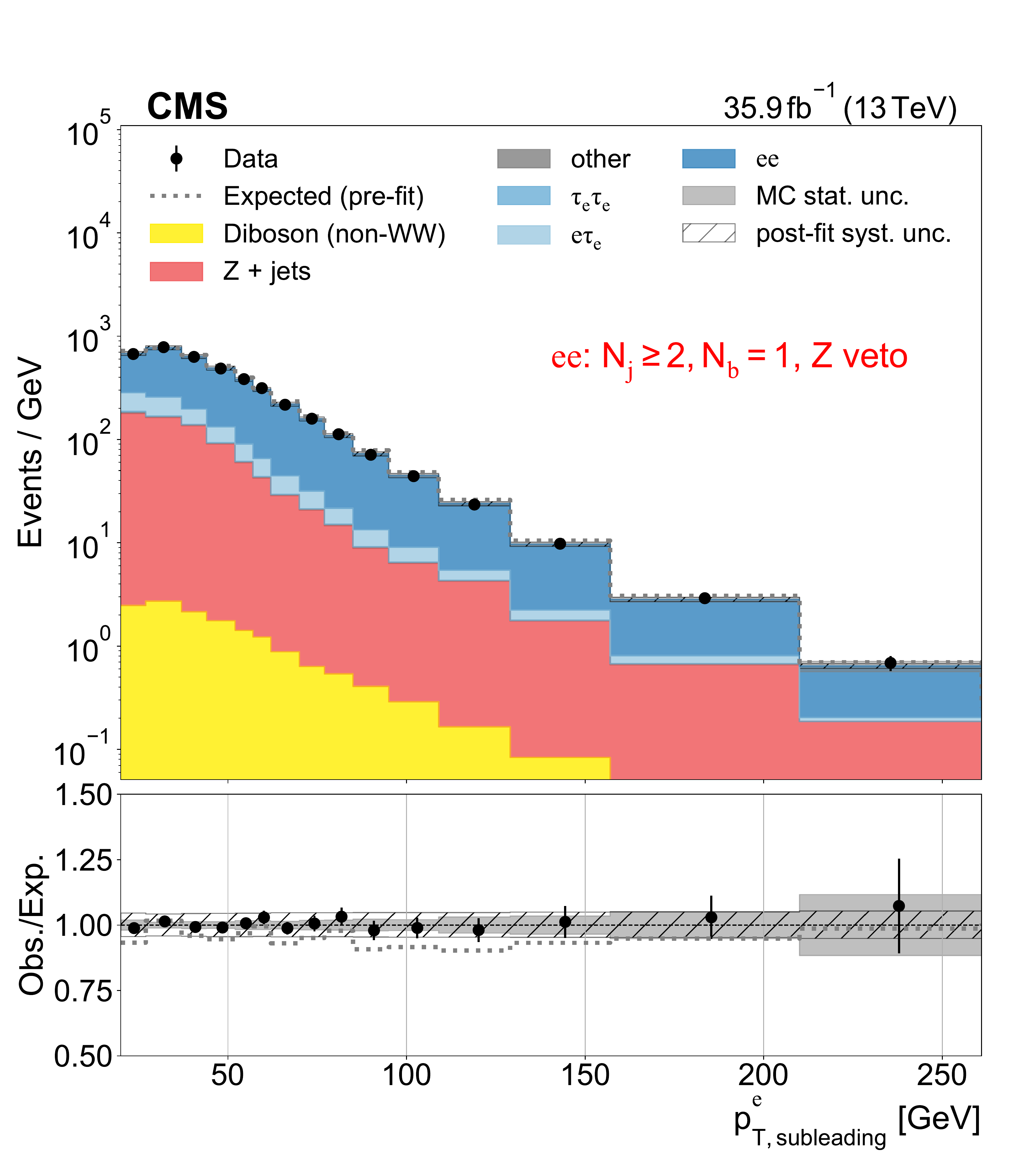}
    \includegraphics[width=0.49\textwidth]{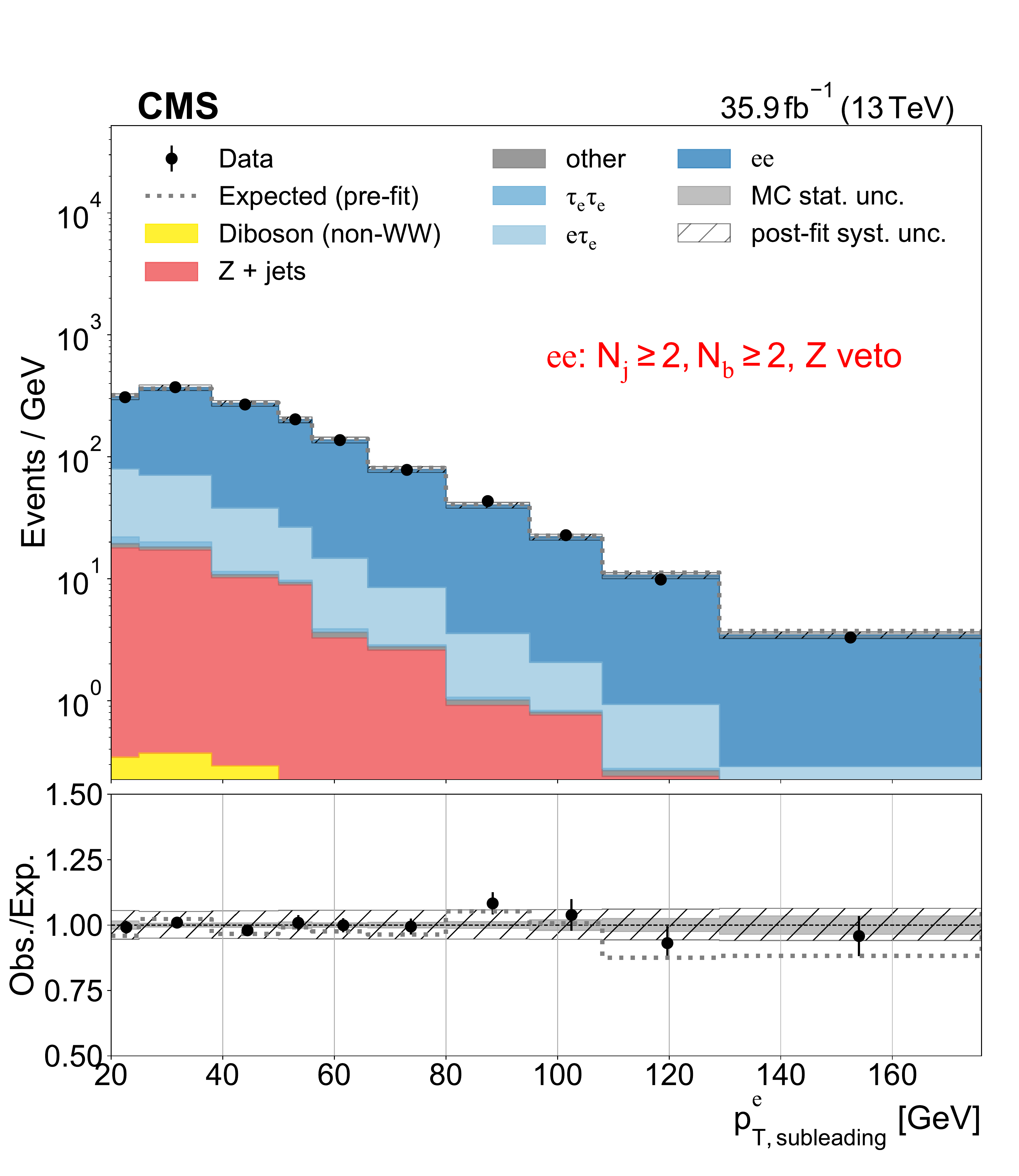}

    \centering
    \includegraphics[width=0.49\textwidth]{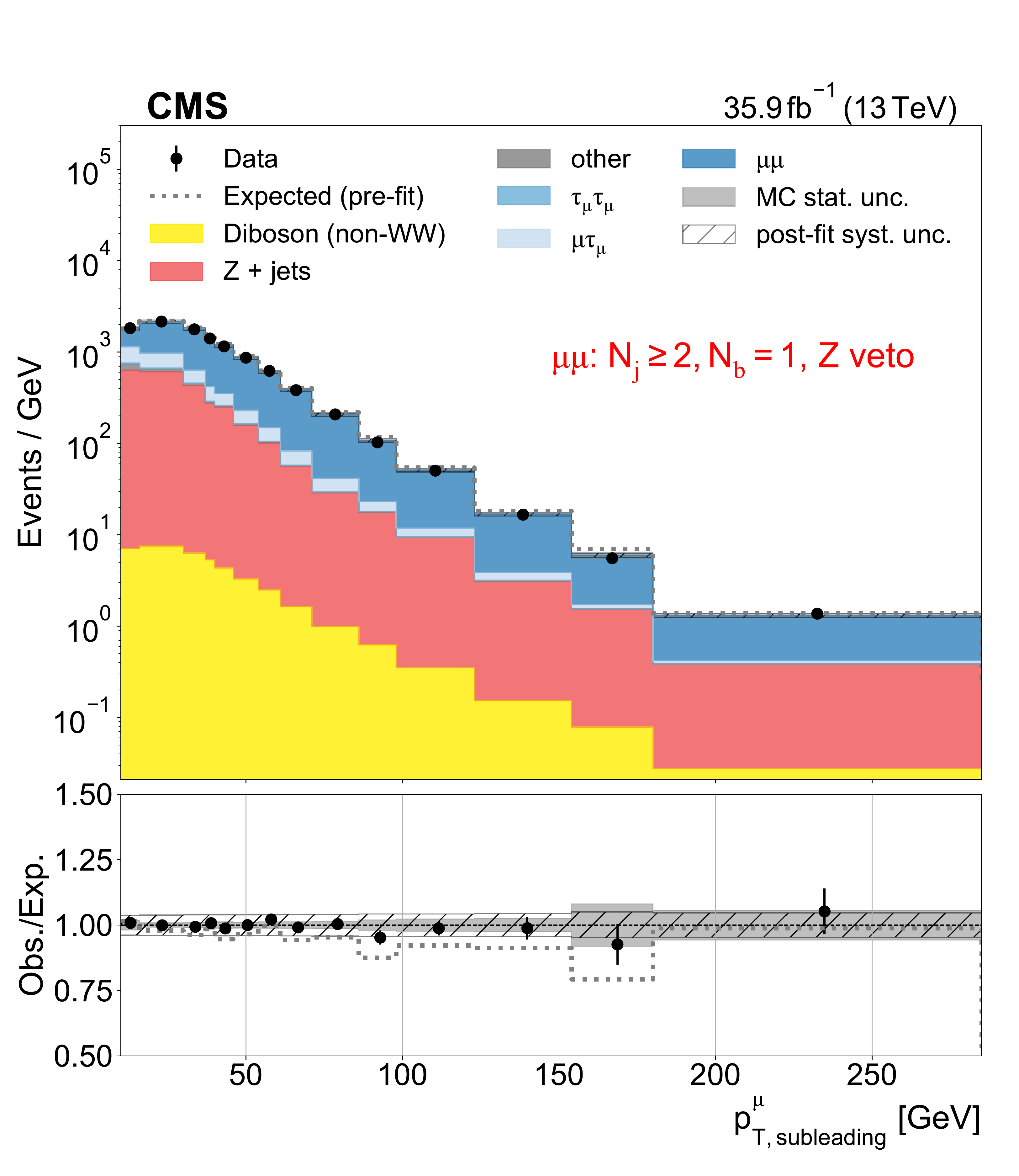}
    \includegraphics[width=0.49\textwidth]{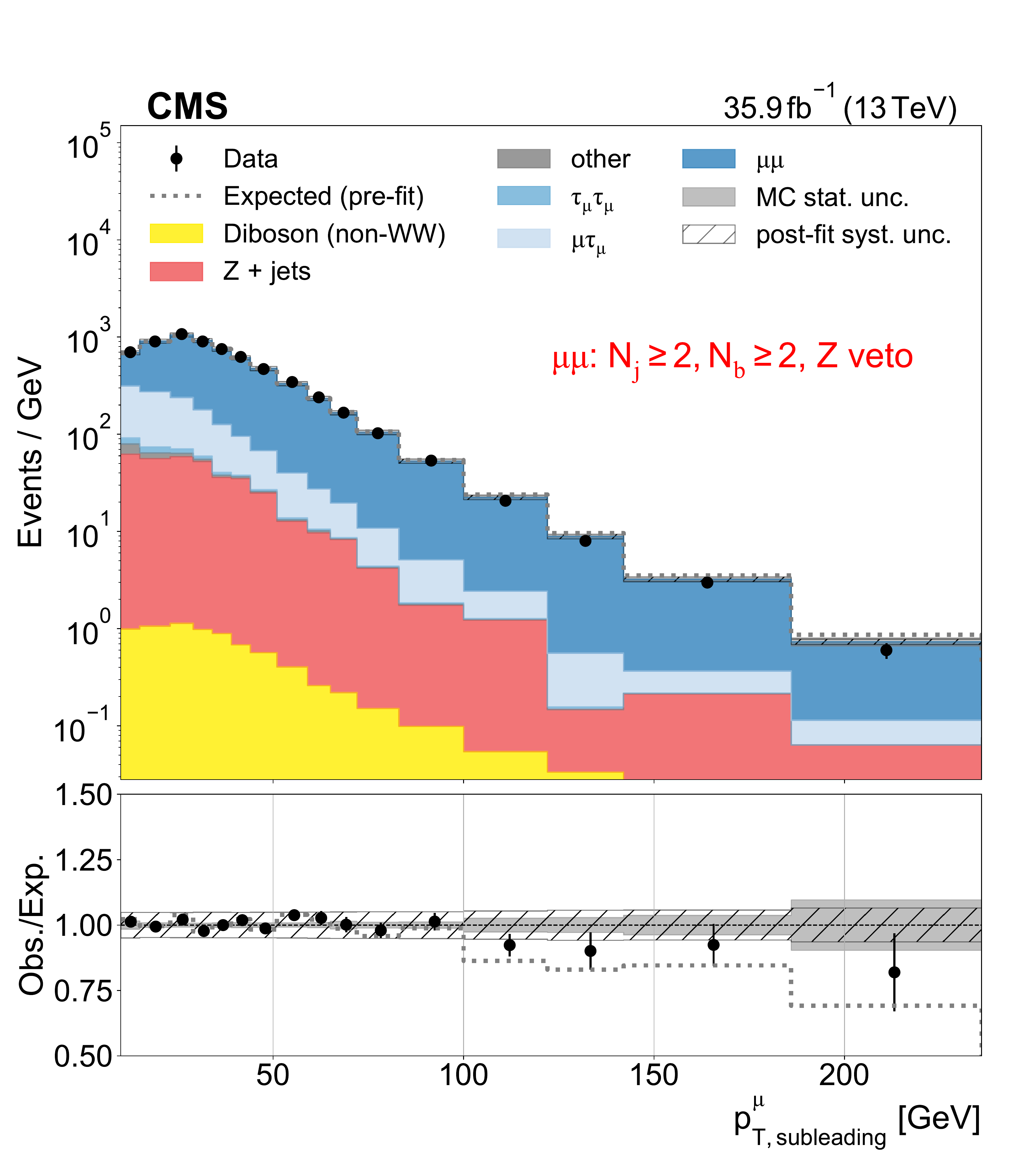}

    \caption{Subleading electron and muon \pt distributions used as inputs for the binned likelihood fits for the $\Pe\Pe$ (upper) and $\PGm\PGm$ (lower) categories, respectively, with the requirement of one (left) or more than one (right) \PQb-tagged jets. The lower subpanels show the ratio of data over pre-fit (dotted line) and post-fit (black circles) expectations,  with associated MC statistical uncertainties (hatched area) and  post-fit systematic uncertainties (shaded gray).  Vertical bars on the data markers indicate statistical uncertainties.
    \label{fig:fits_templates_ll}}
\end{figure*}

\begin{figure*}[htbp!]
    \centering
    \includegraphics[width=0.35\textwidth]{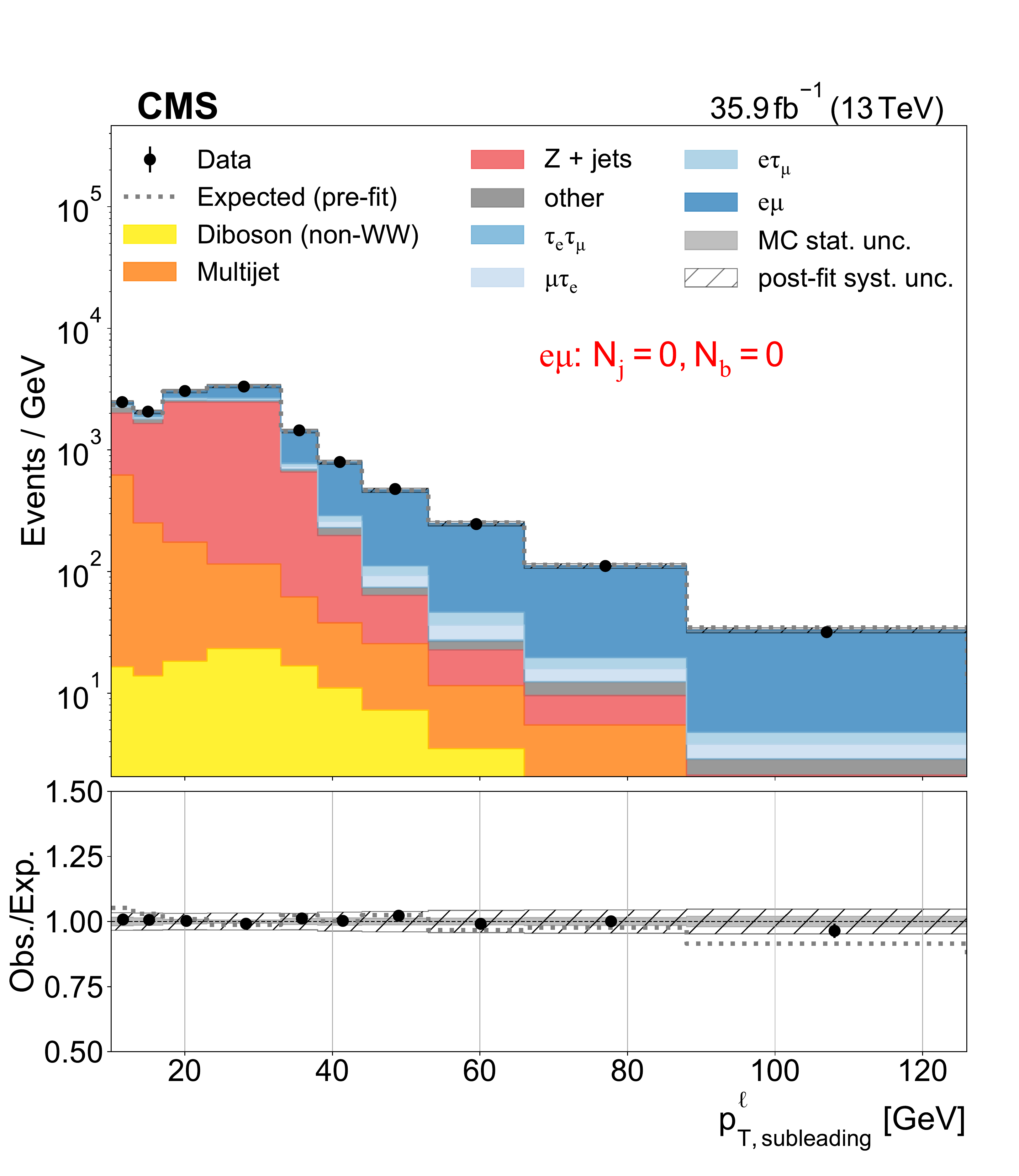}
    \includegraphics[width=0.35\textwidth]{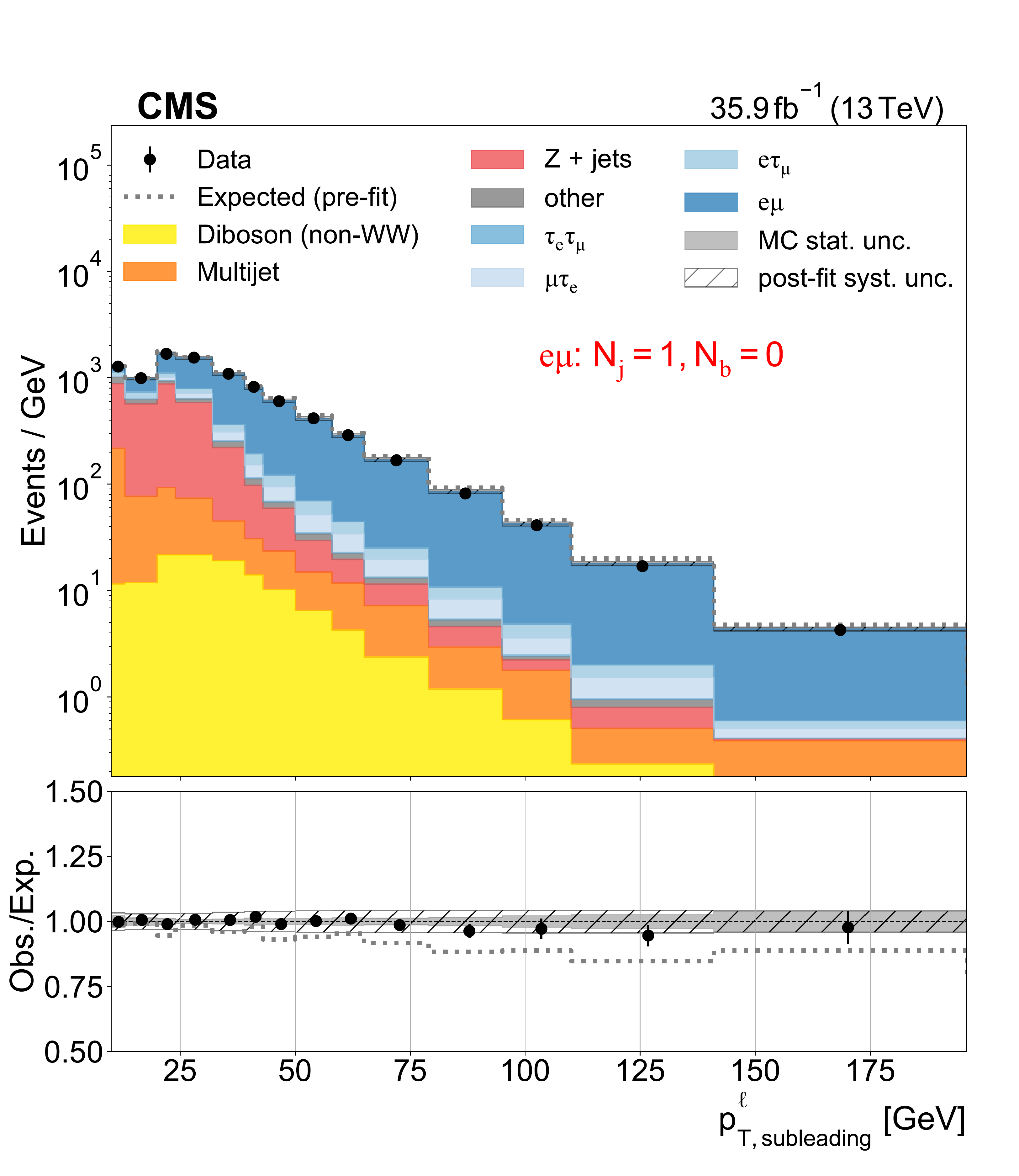}

    \includegraphics[width=0.35\textwidth]{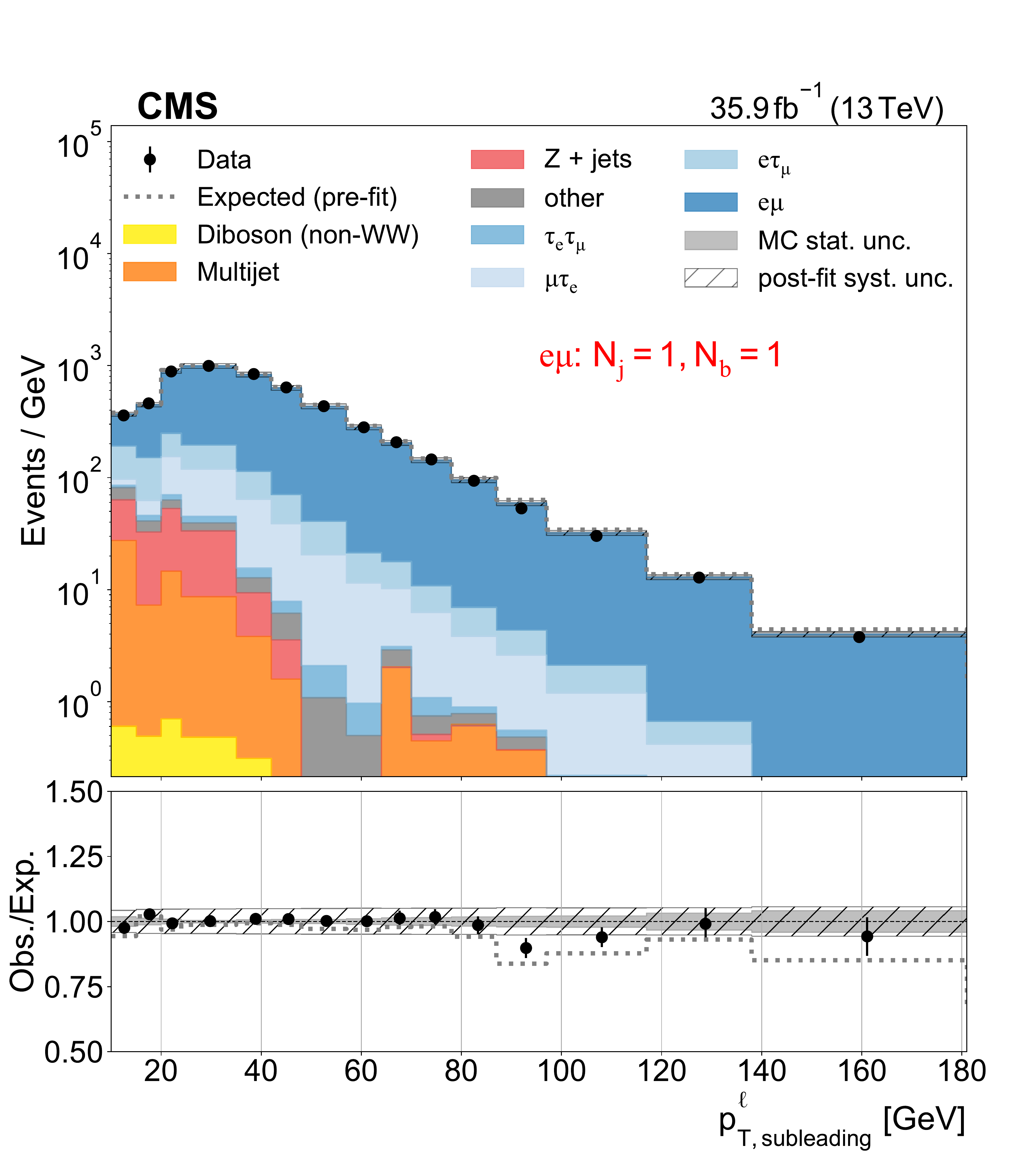}
    \includegraphics[width=0.35\textwidth]{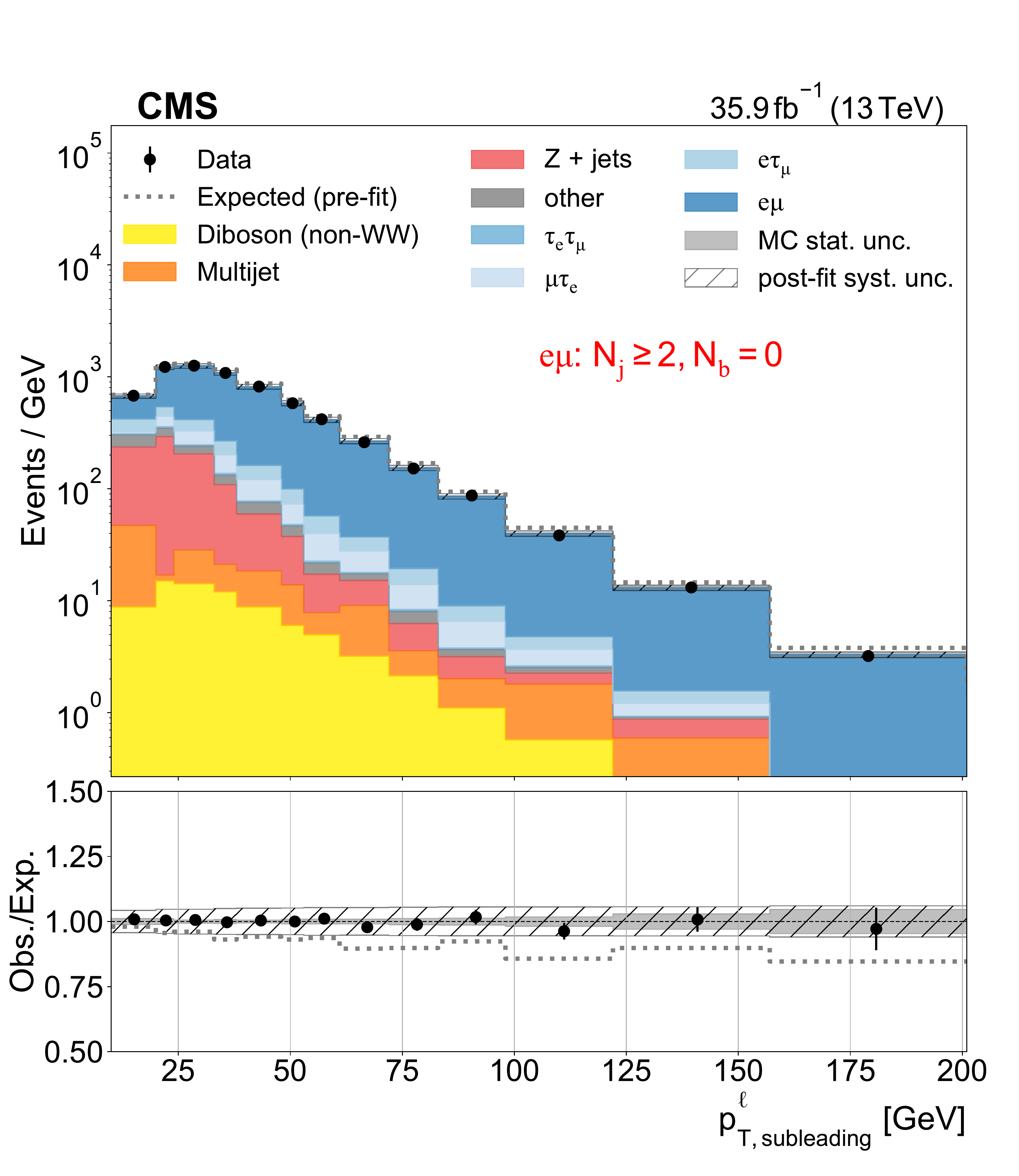}

    \includegraphics[width=0.35\textwidth]{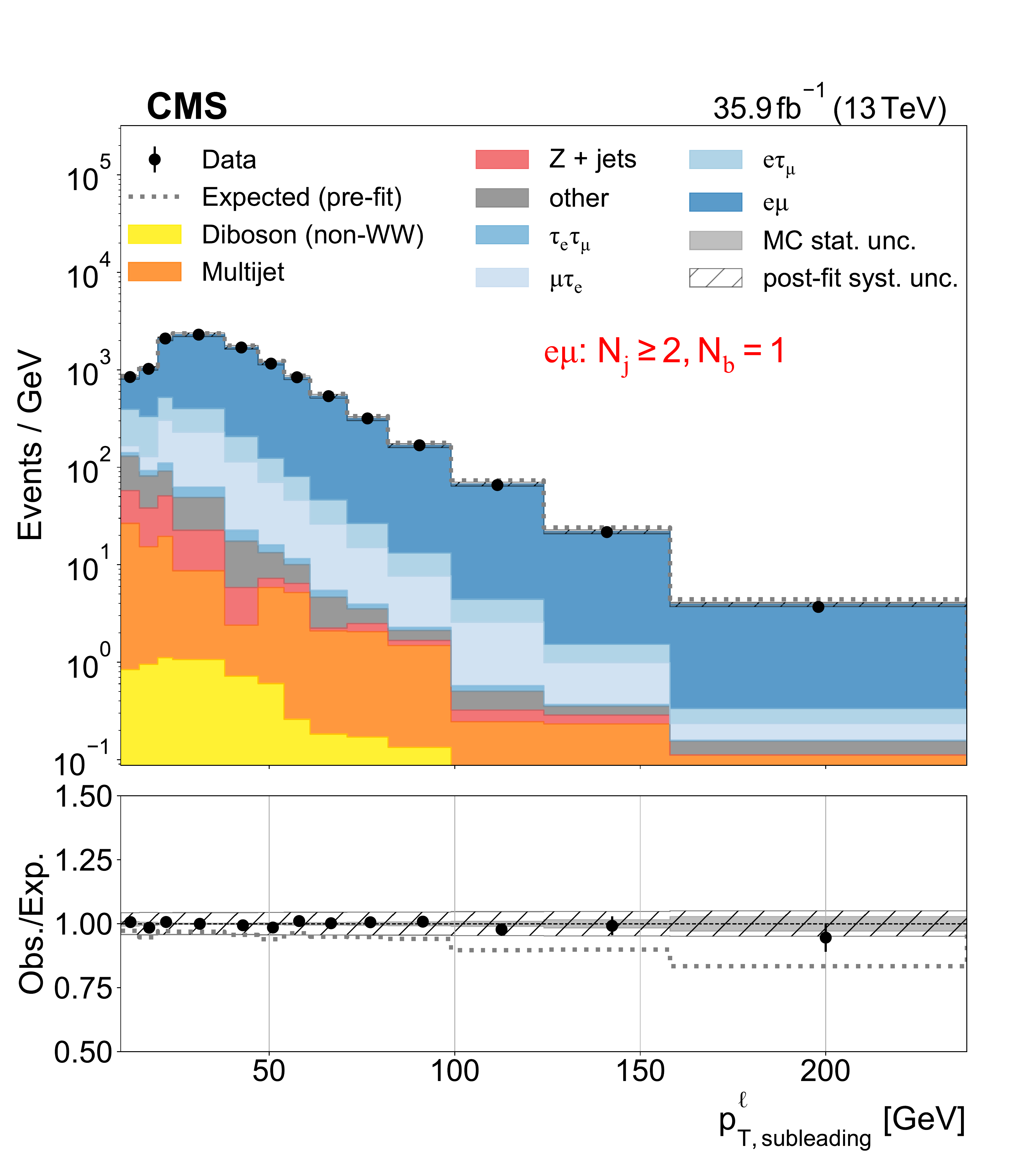}
    \includegraphics[width=0.35\textwidth]{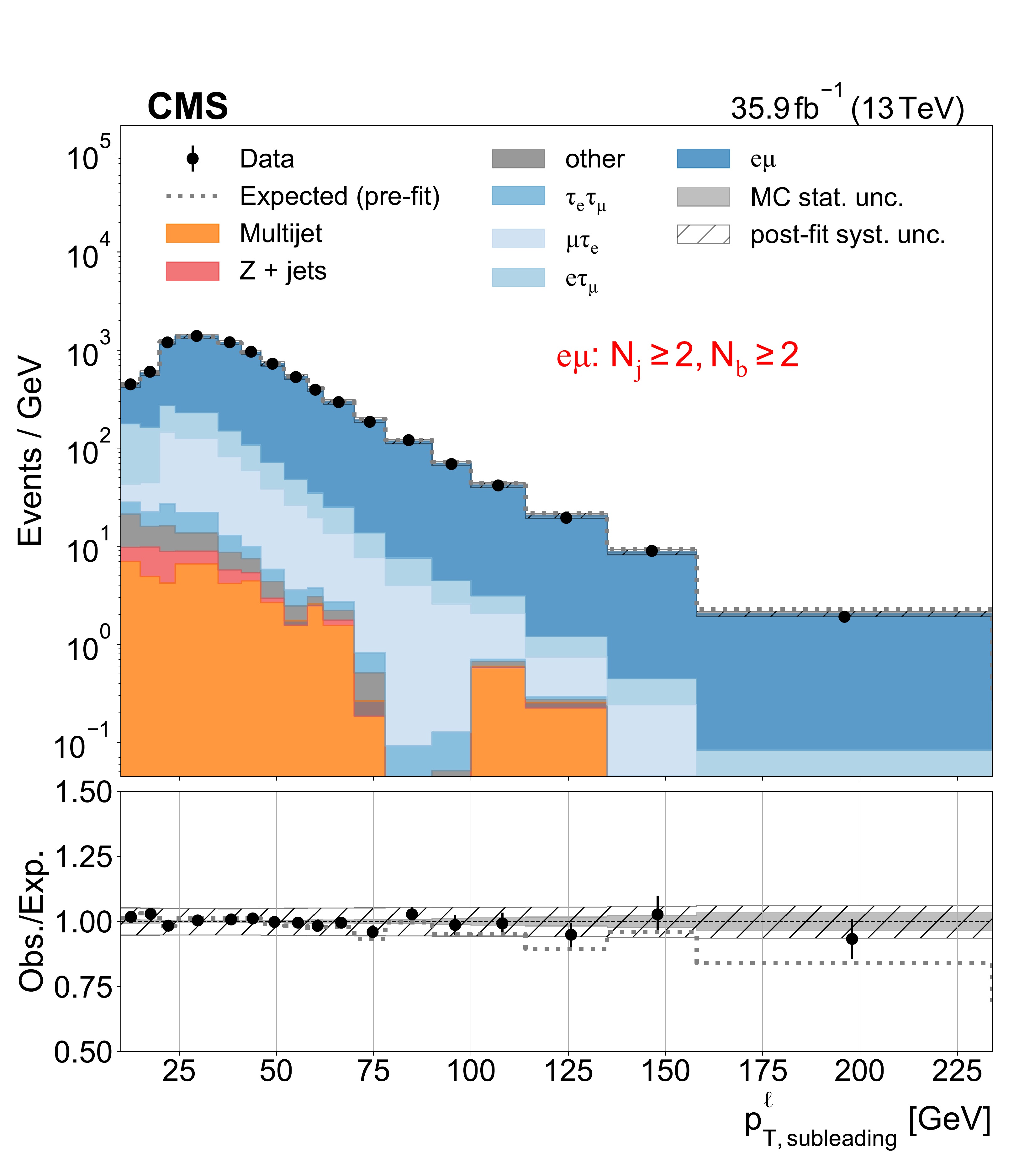}
    \caption{Subleading lepton, electron or muon, \pt distributions used as inputs for the binned likelihood fits for the $\Pe\PGm$ categories. The different panels are obtained with the listed selection criteria on the number of jets ($N_{\jet}$) and of \PQb-tagged jets ($N_{\PQb}$) required. The lower subpanels show the ratio of data over pre-fit expectations, with the gray histograms (hatched area) indicating MC statistical (post-fit systematic) uncertainties. Vertical bars on the data markers indicate statistical uncertainties.
    \label{fig:fits_templates_emu}}
\end{figure*}

\begin{figure*}[htbp!]
    \centering
    \includegraphics[width=0.325\textwidth]{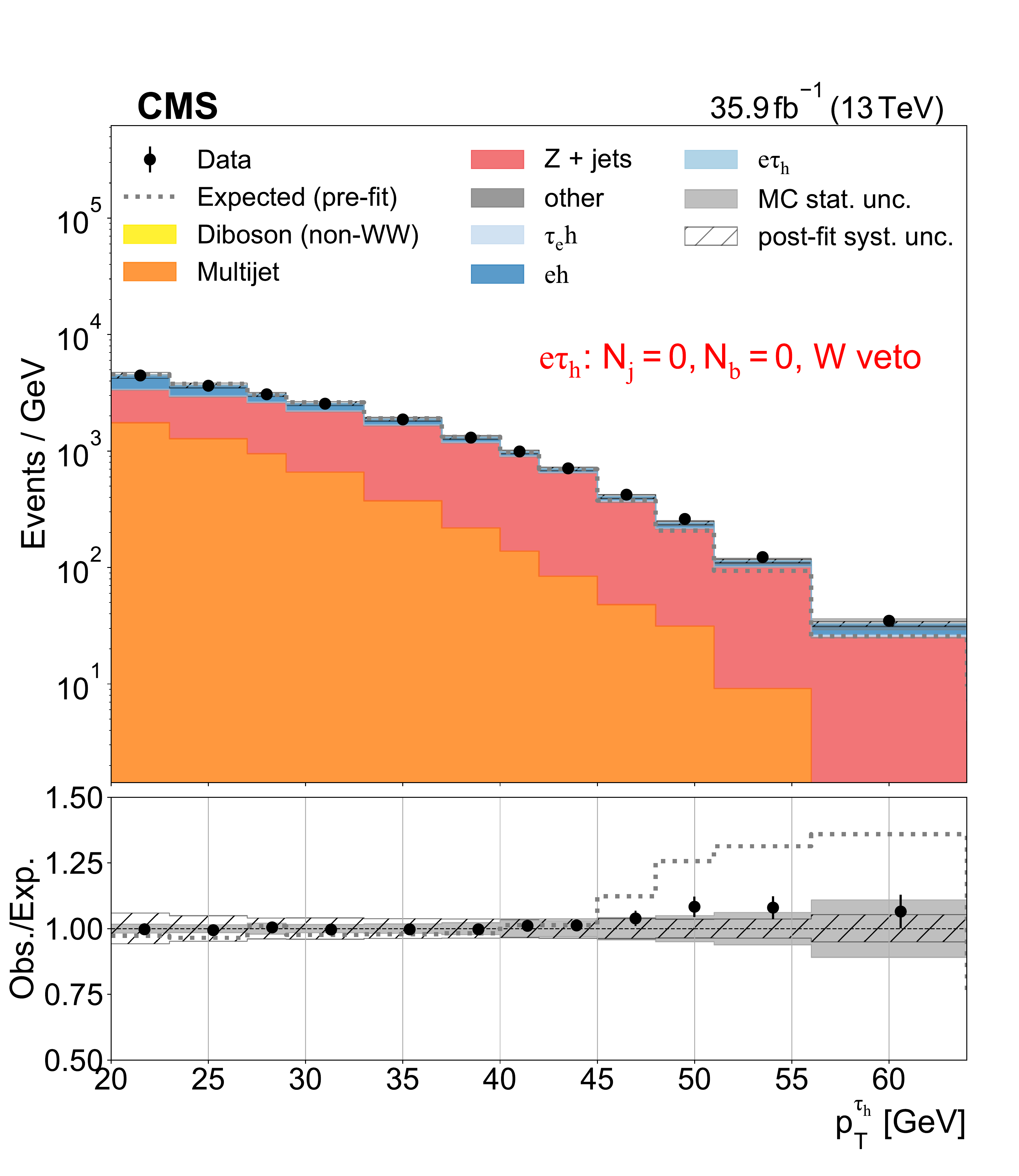}
    \includegraphics[width=0.325\textwidth]{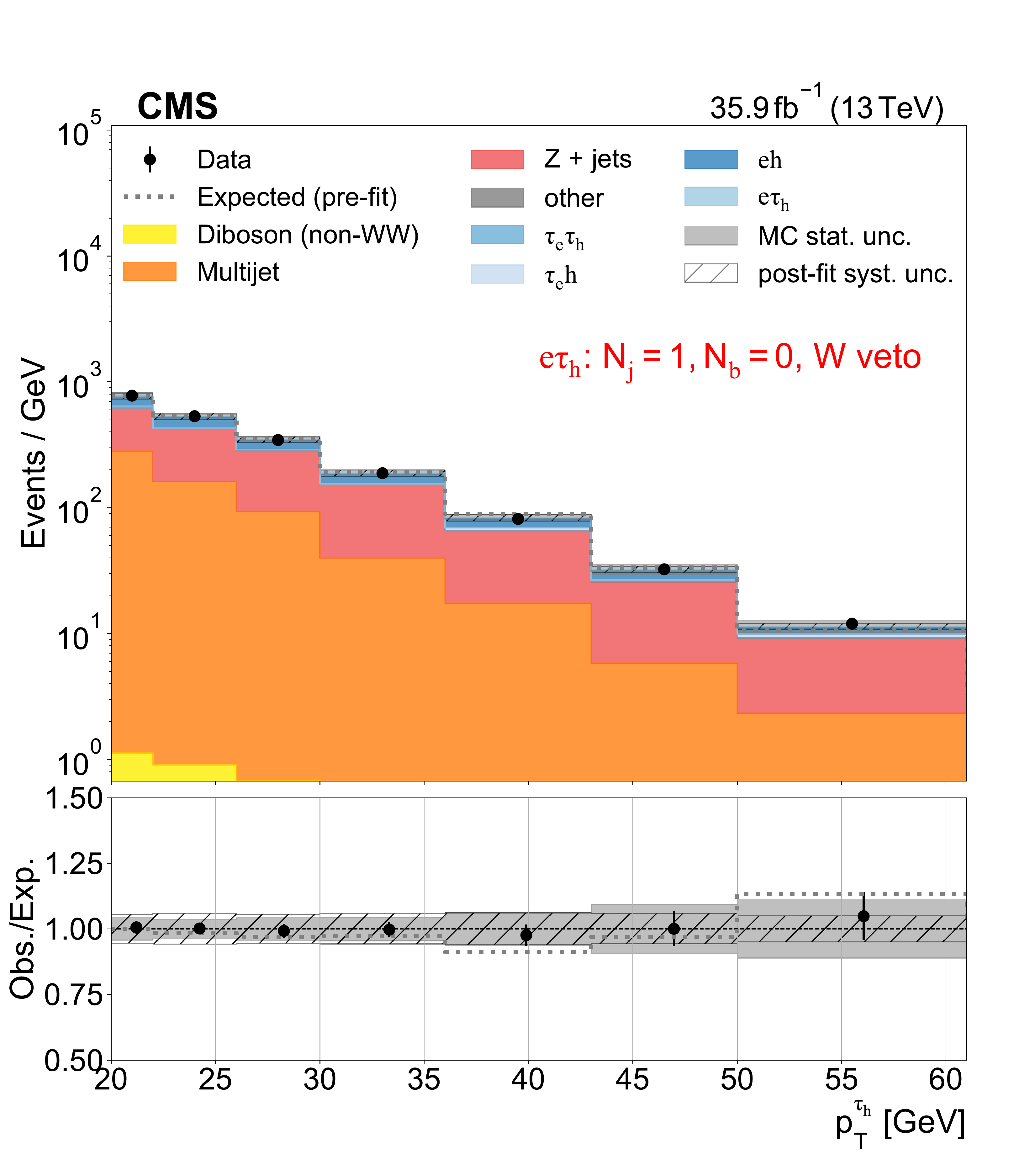}
    \includegraphics[width=0.325\textwidth]{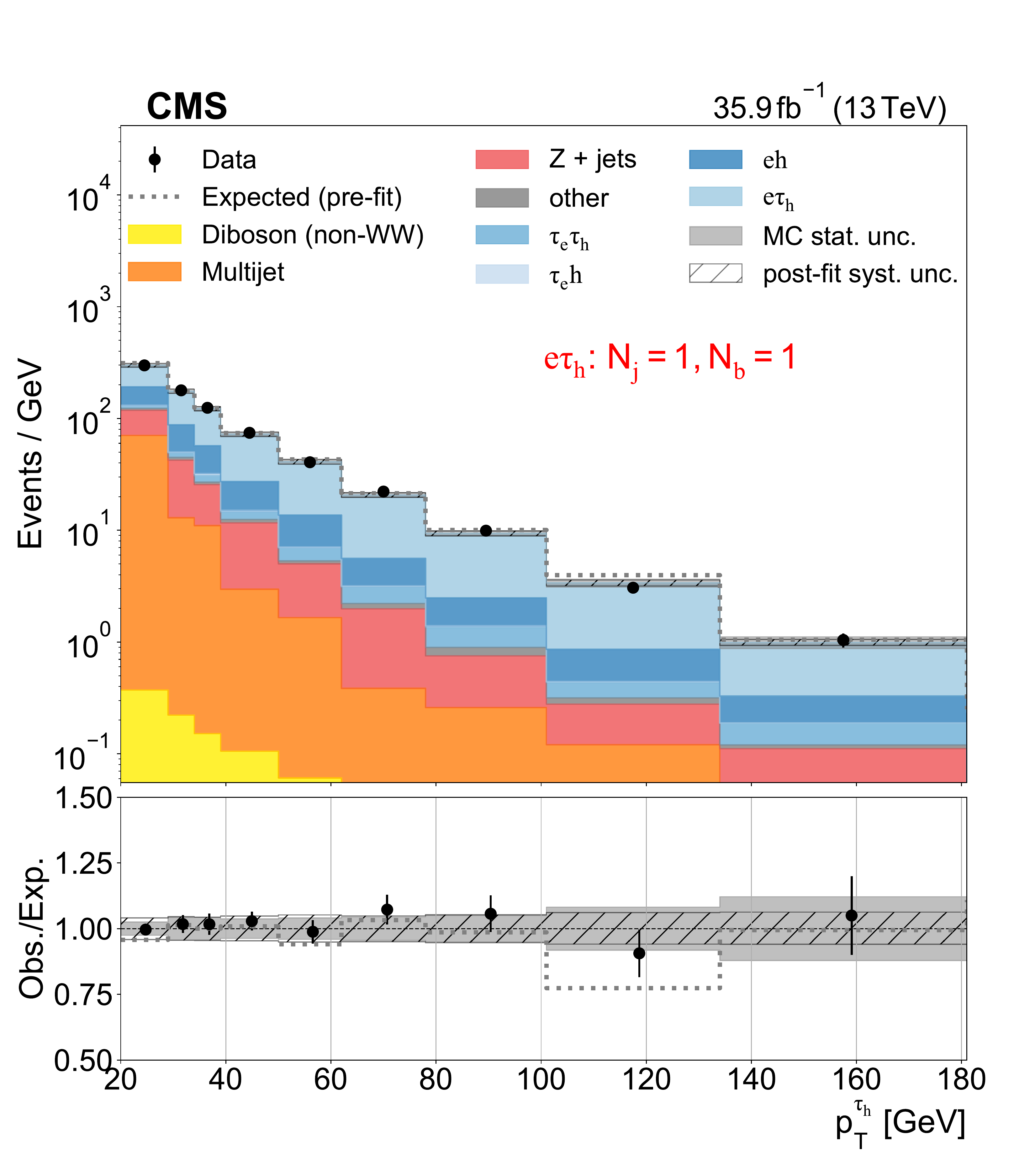}
    
    \includegraphics[width=0.325\textwidth]{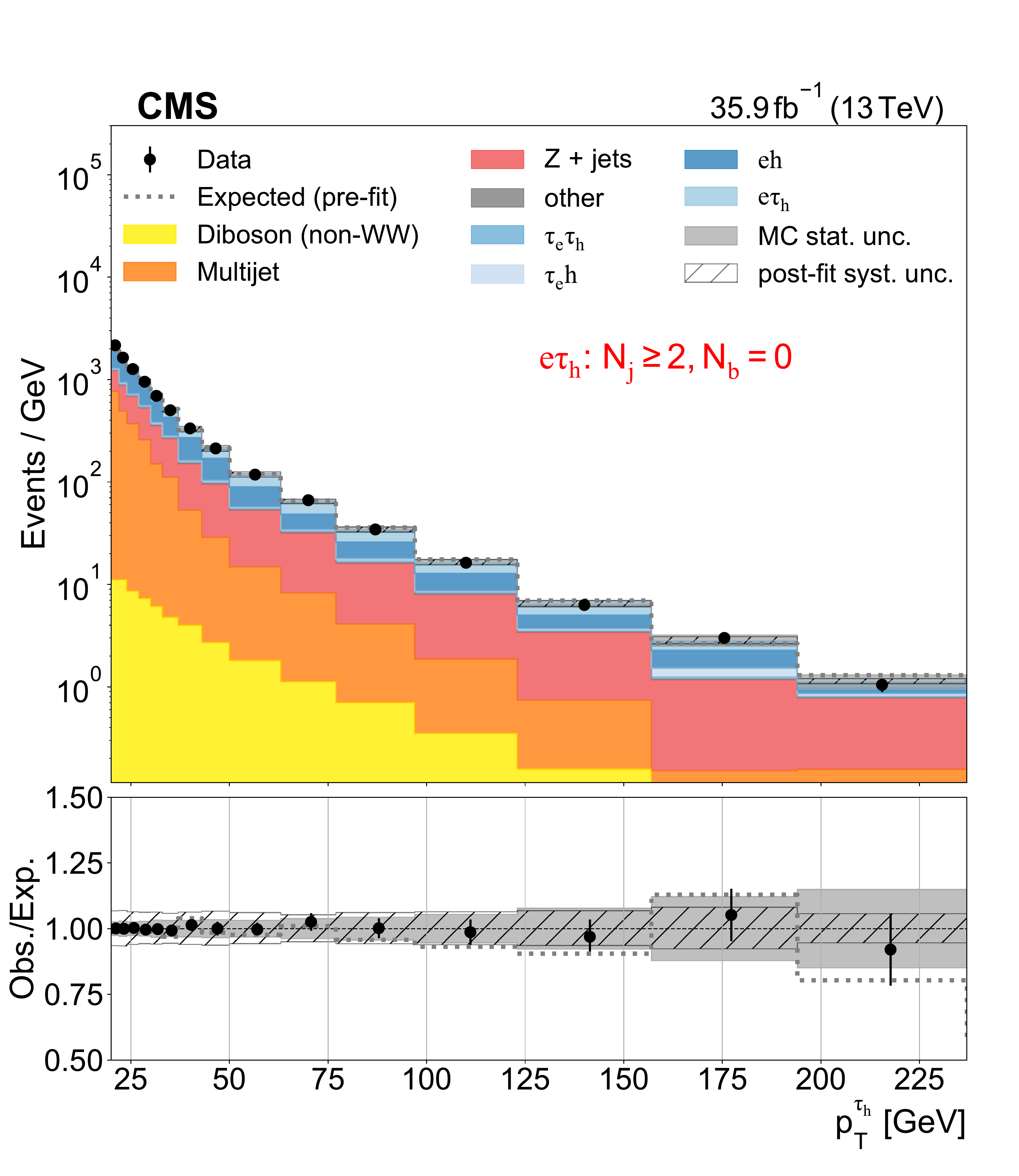}
    \includegraphics[width=0.325\textwidth]{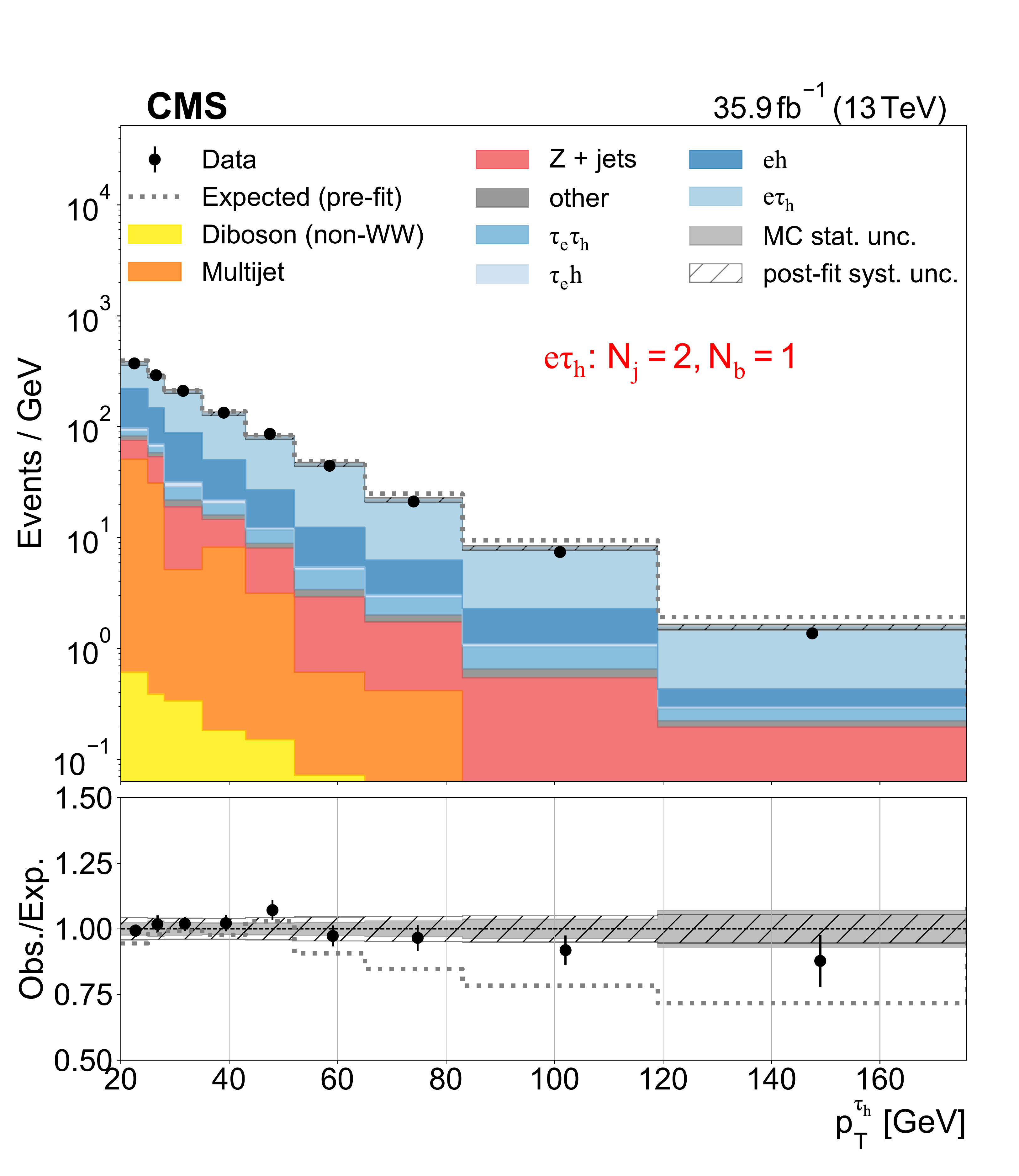}
    \includegraphics[width=0.325\textwidth]{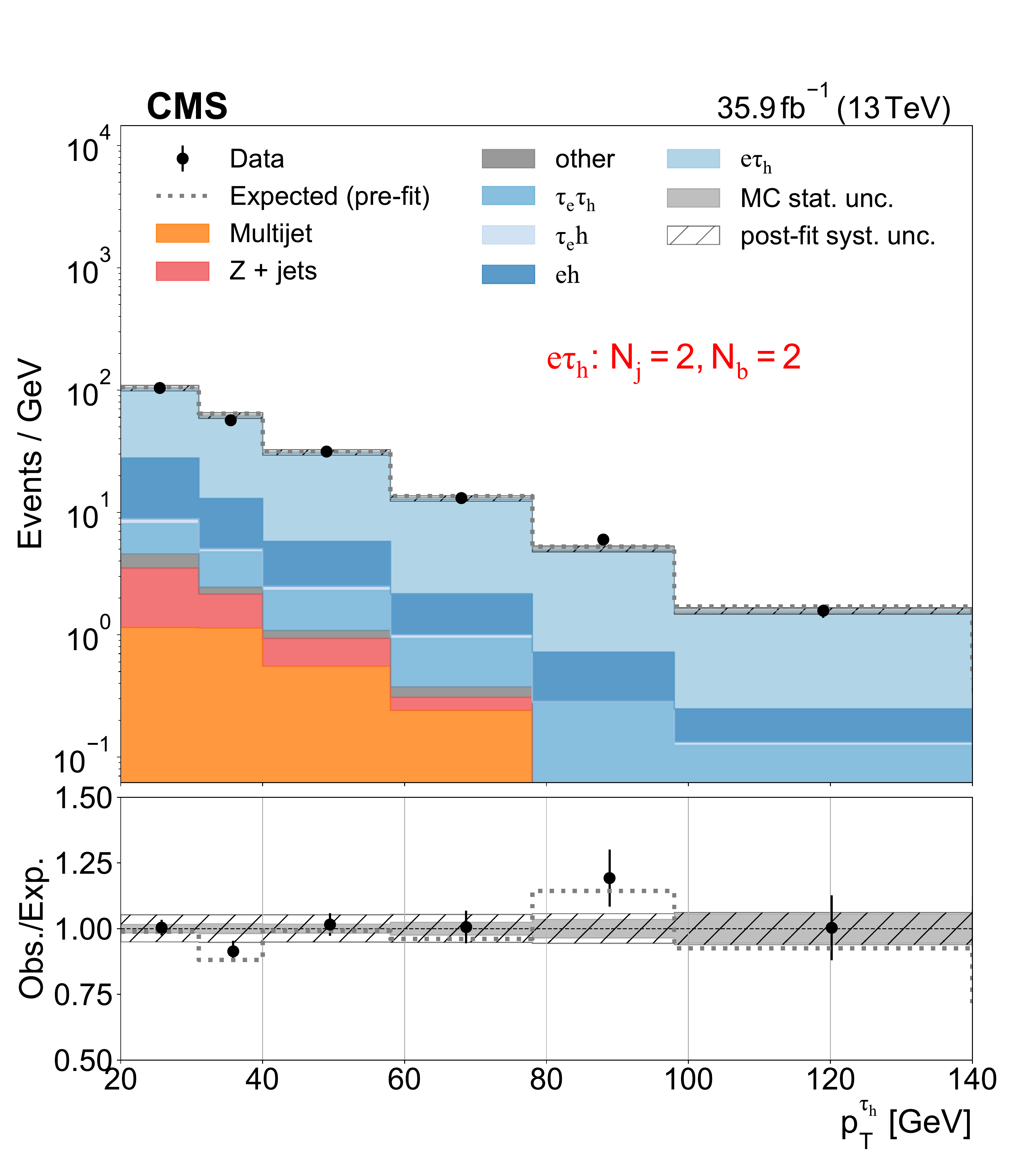}
    
    \includegraphics[width=0.325\textwidth]{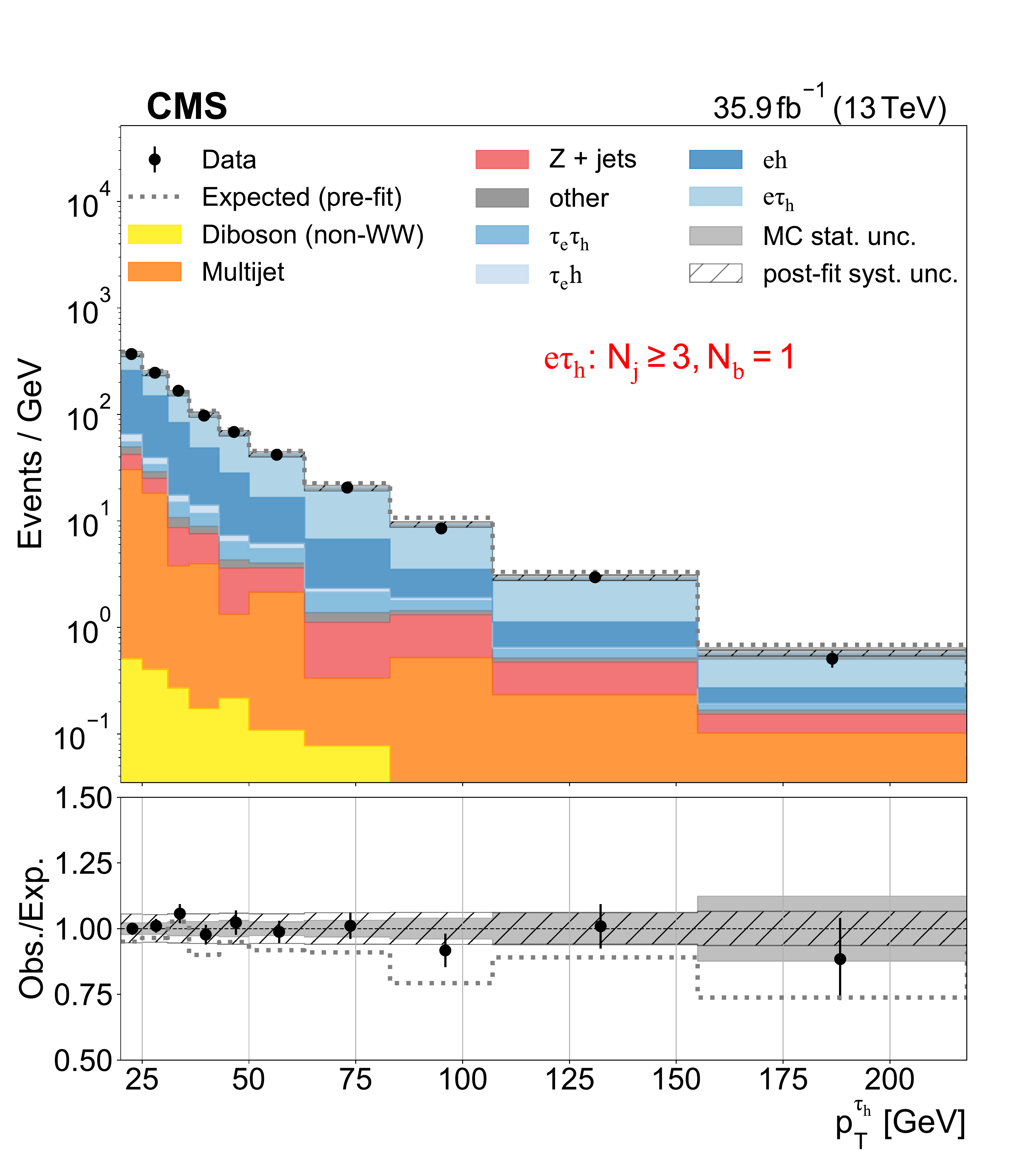}
    \includegraphics[width=0.325\textwidth]{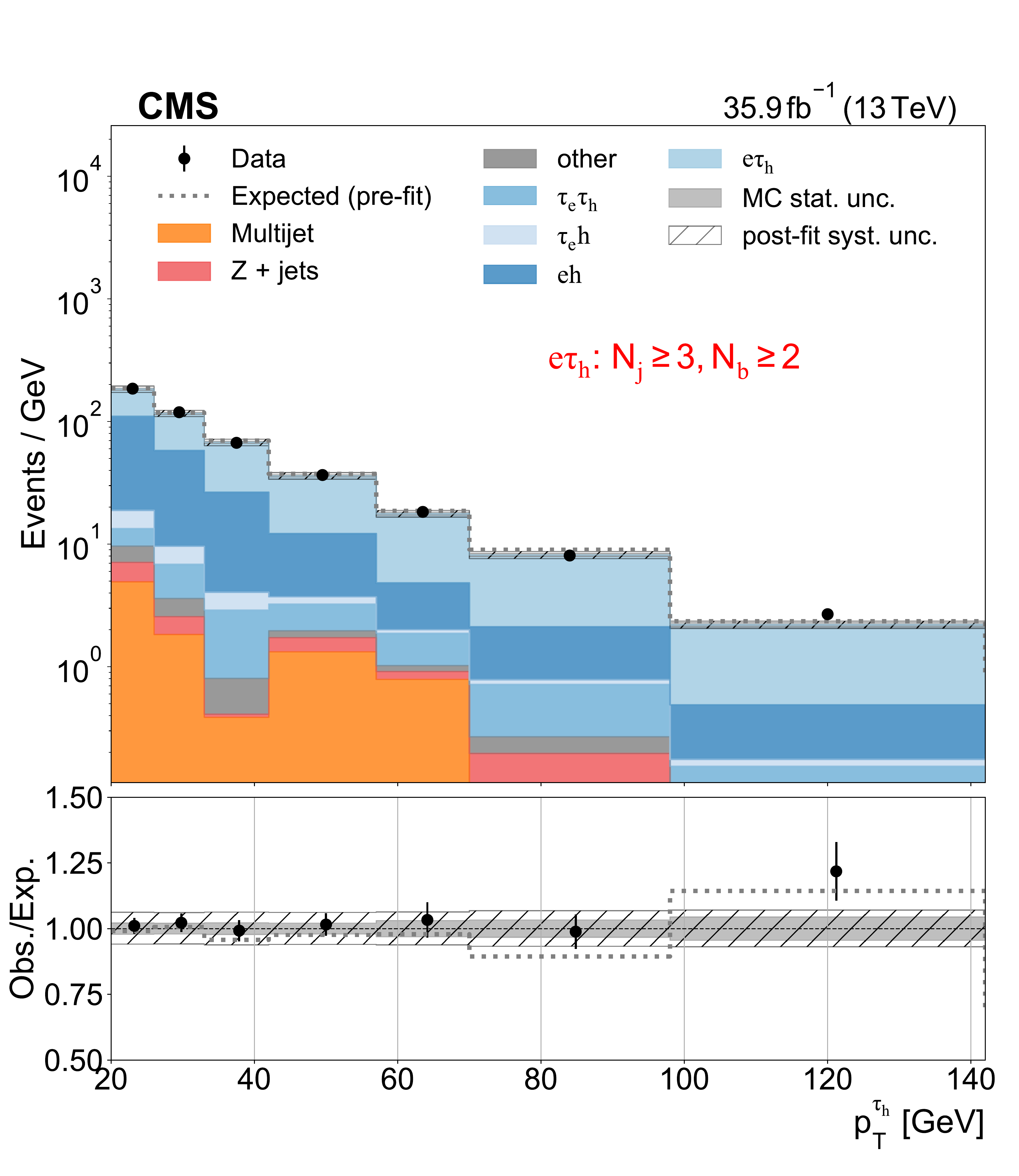}
    \caption{Distributions of \tauh \pt used as inputs for the binned likelihood fits for the $\Pe\PGt$ categories. The different panels list the varying selections on the number of jets ($N_{\jet}$) and of \PQb-tagged jets ($N_{\PQb}$) required in each case. The lower subpanels show the ratio of data over pre-fit expectations, with the gray band (hatched area) indicating MC statistical (post-fit systematic) uncertainties. Vertical bars on the data markers indicate statistical uncertainties.
    \label{fig:fits_templates_etau}}
\end{figure*}

\begin{figure*}[htbp!]
    \centering
    \includegraphics[width=0.325\textwidth]{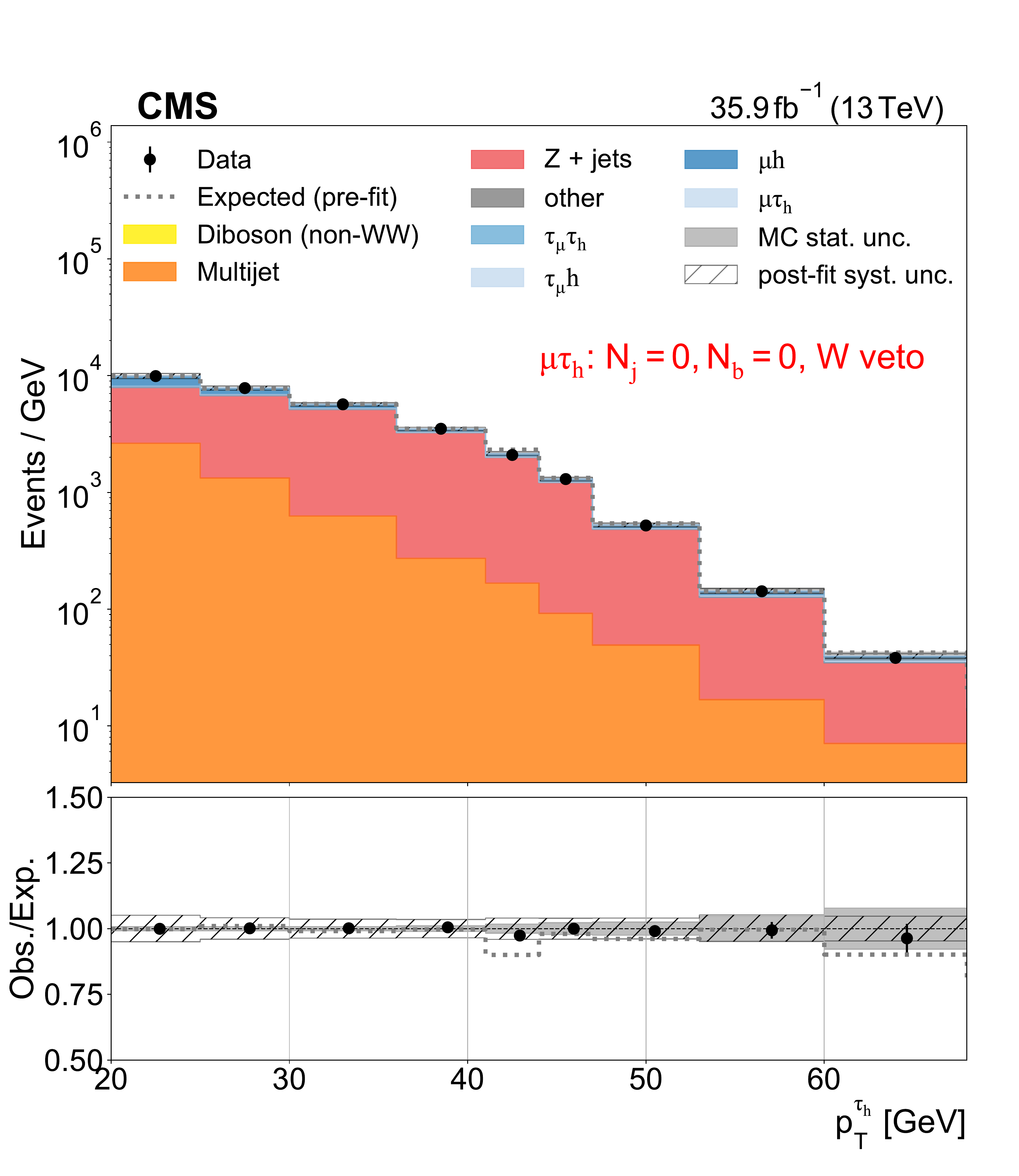}
    \includegraphics[width=0.325\textwidth]{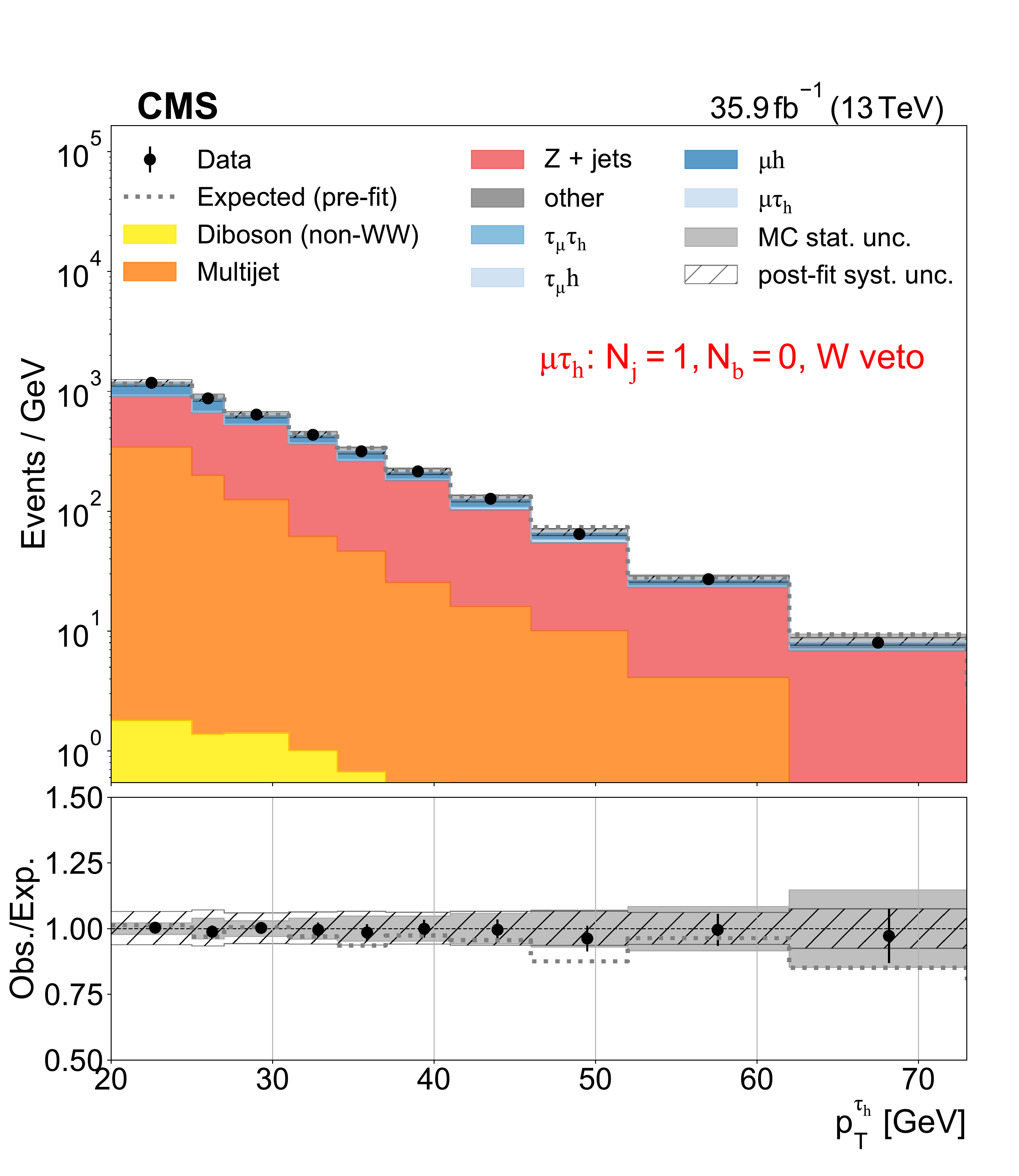}
    \includegraphics[width=0.325\textwidth]{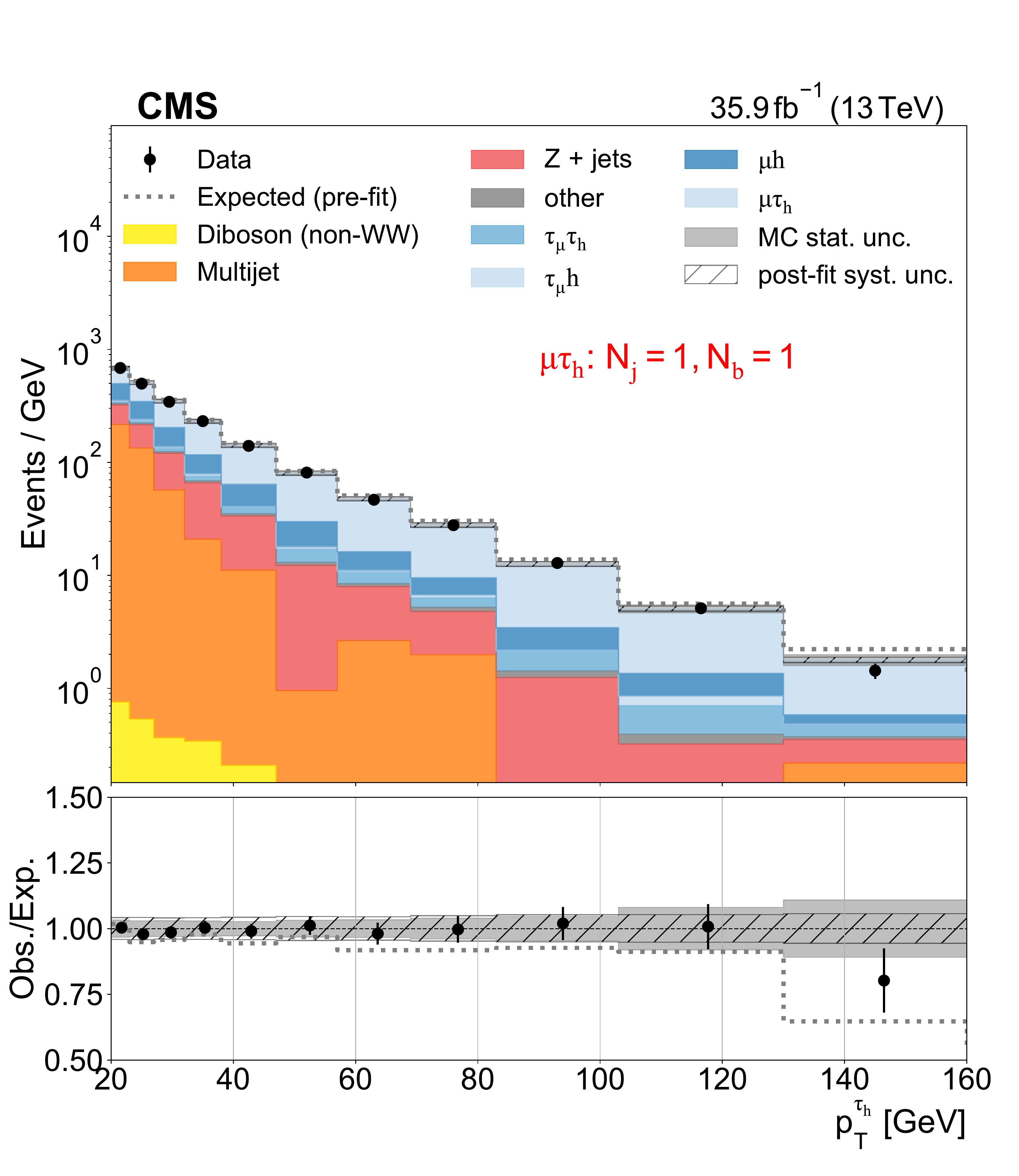}
    
    \includegraphics[width=0.325\textwidth]{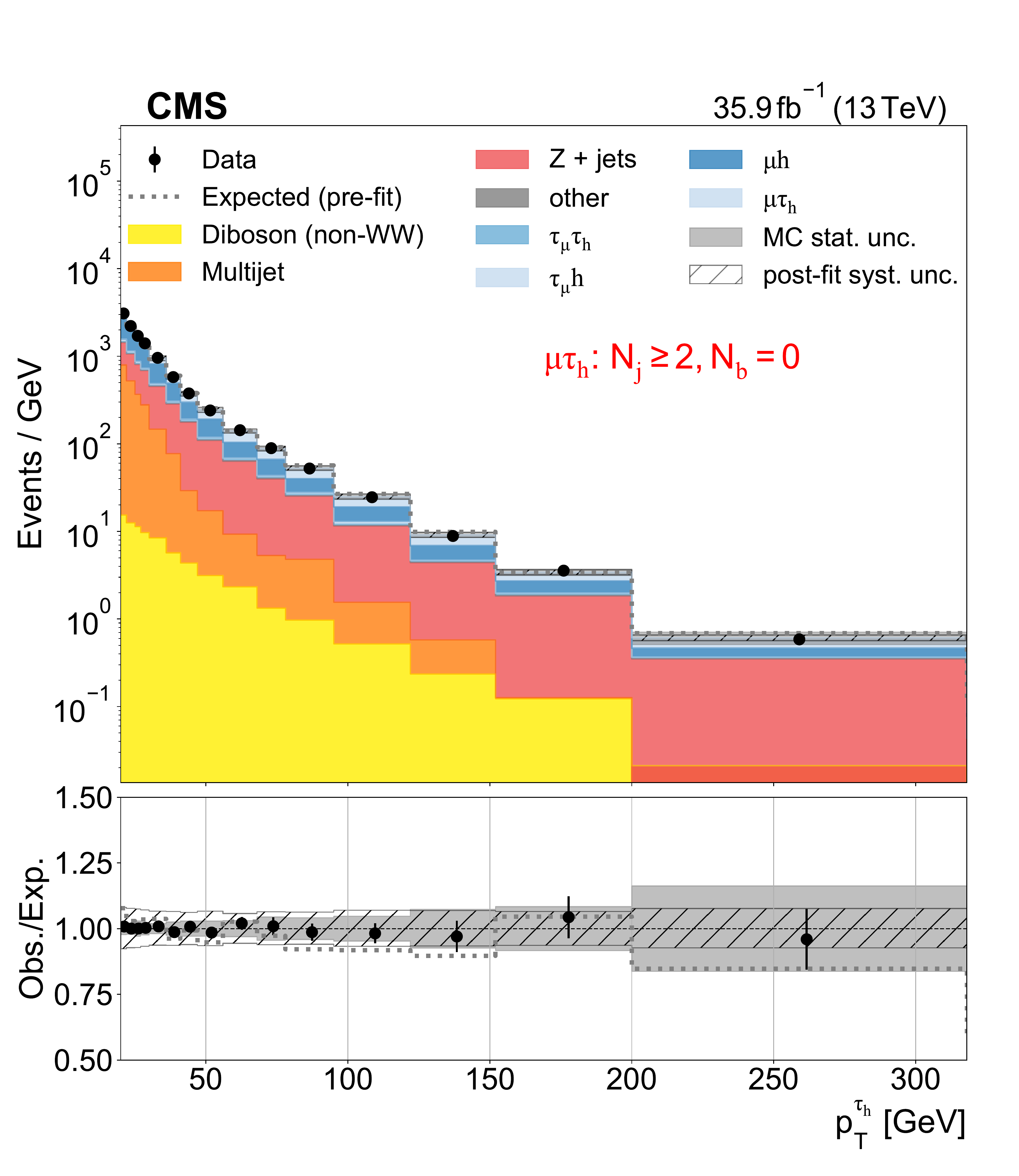}
    \includegraphics[width=0.325\textwidth]{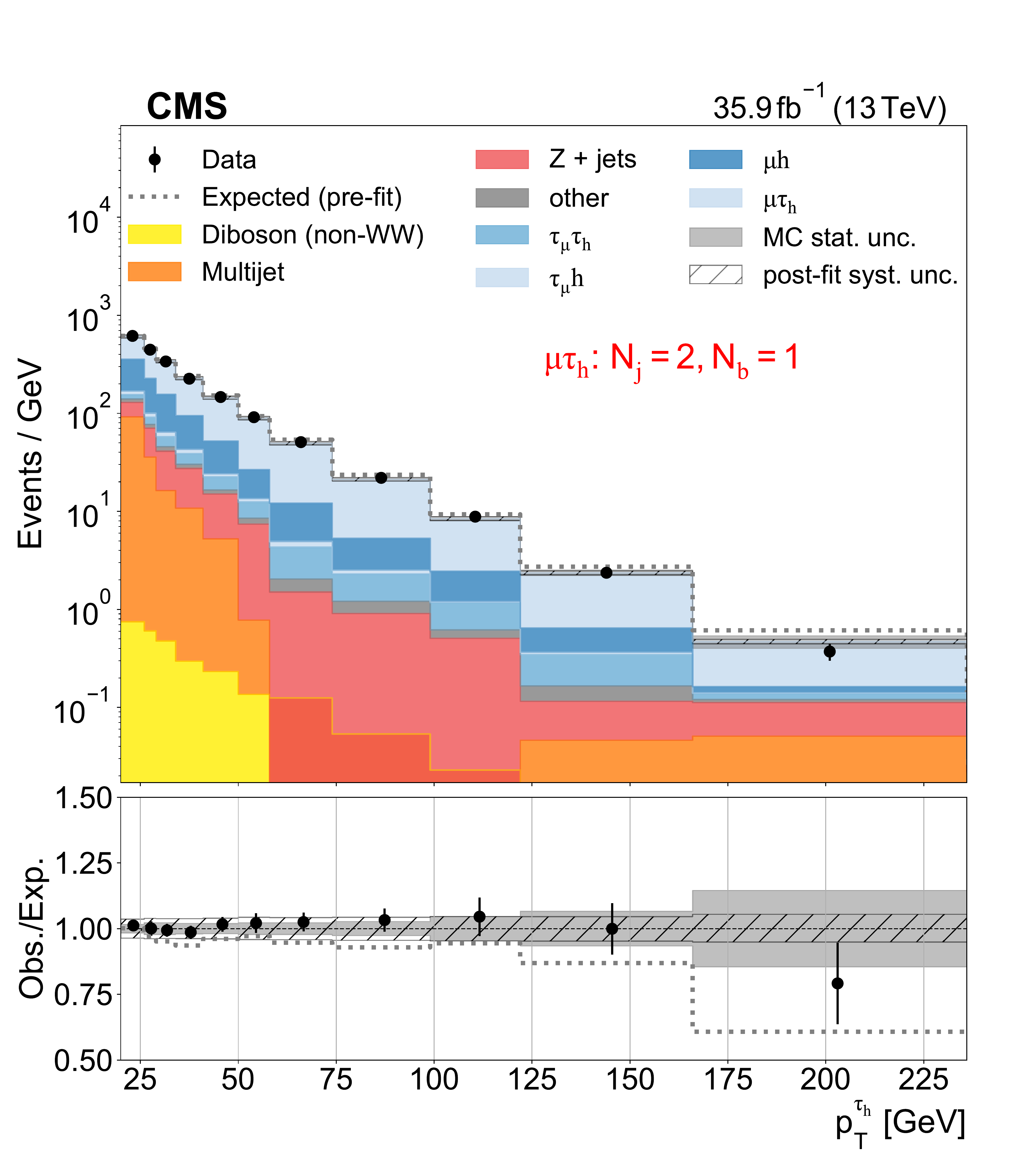}
    \includegraphics[width=0.325\textwidth]{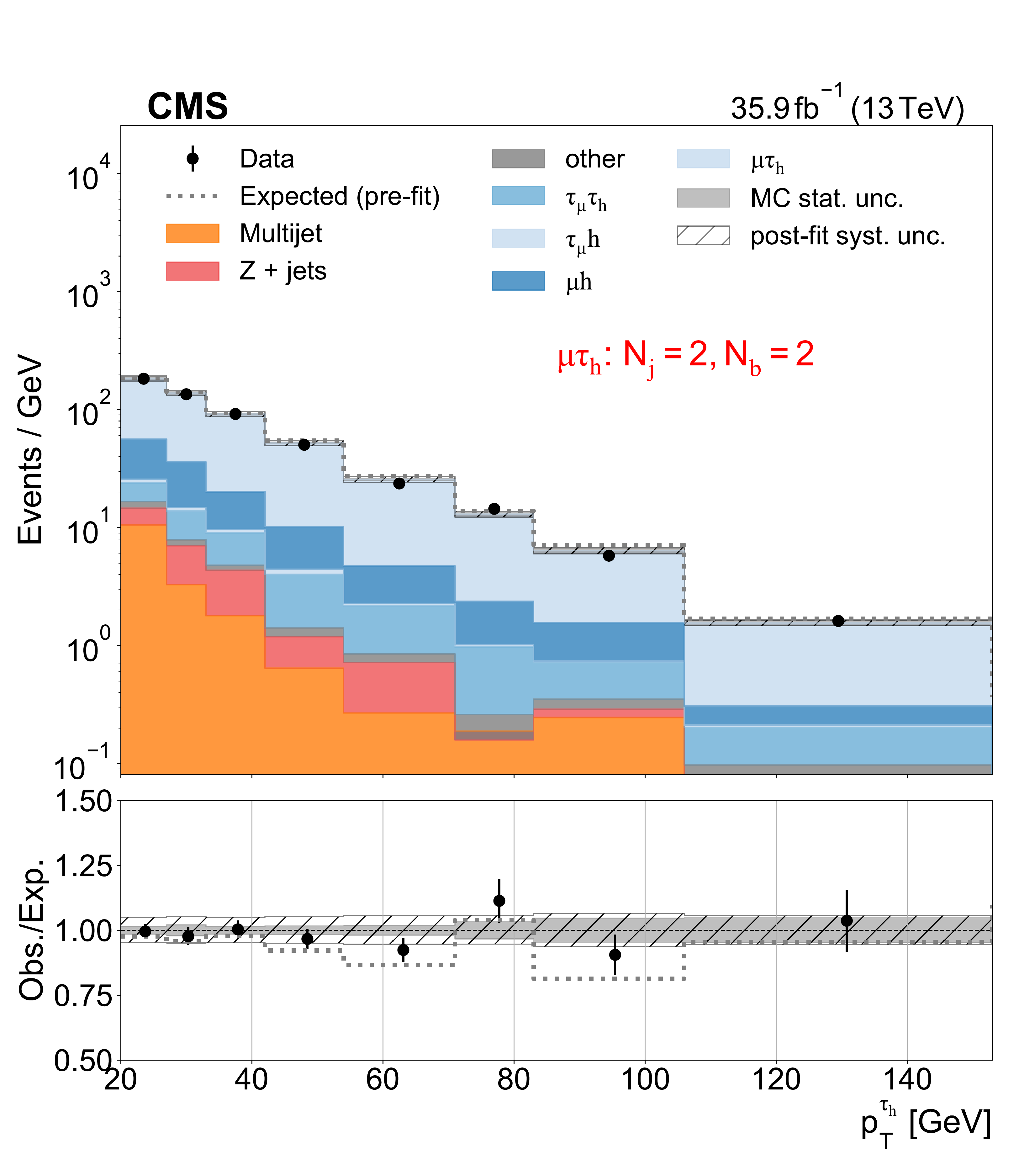}
    
    \includegraphics[width=0.325\textwidth]{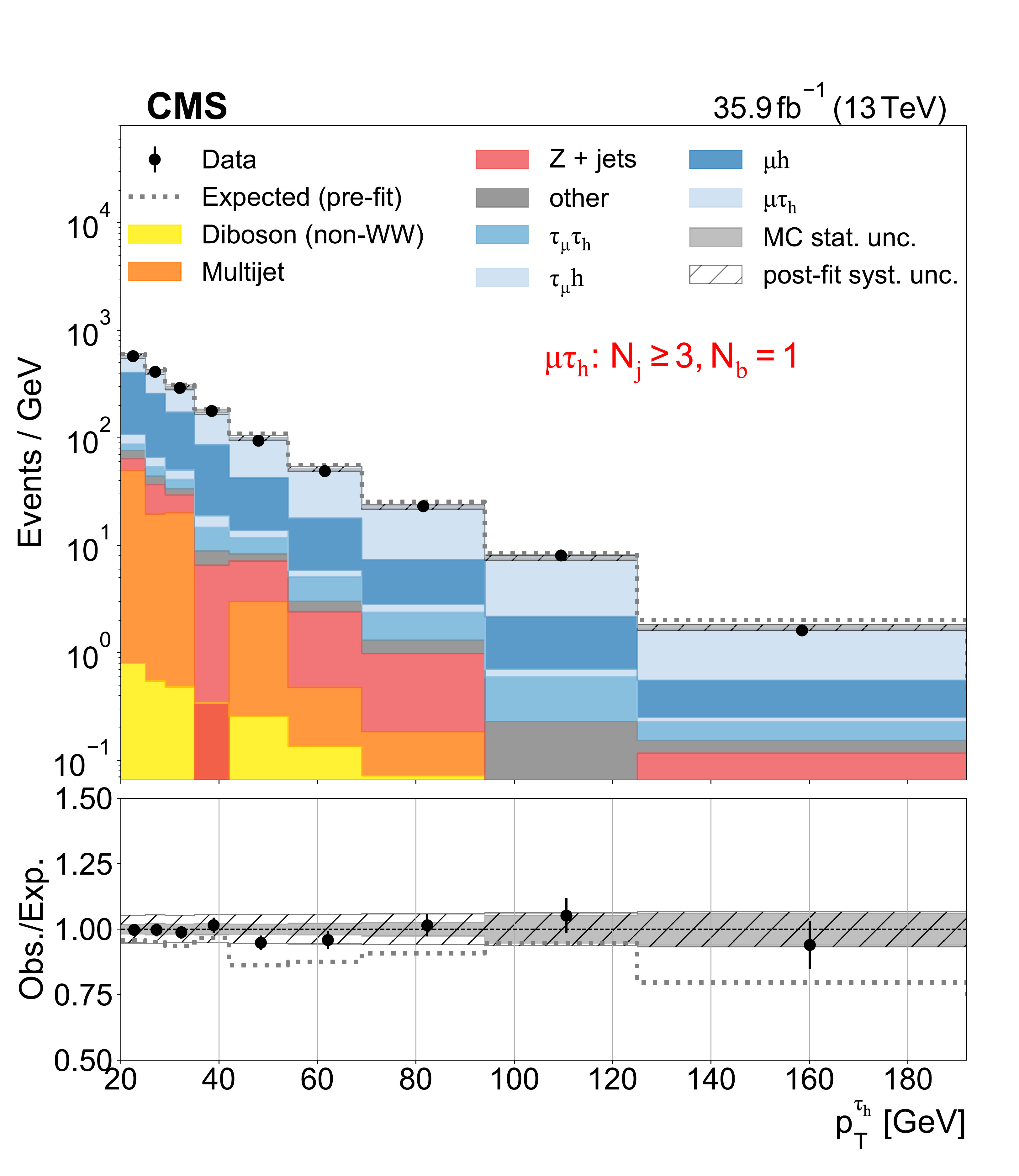}
    \includegraphics[width=0.325\textwidth]{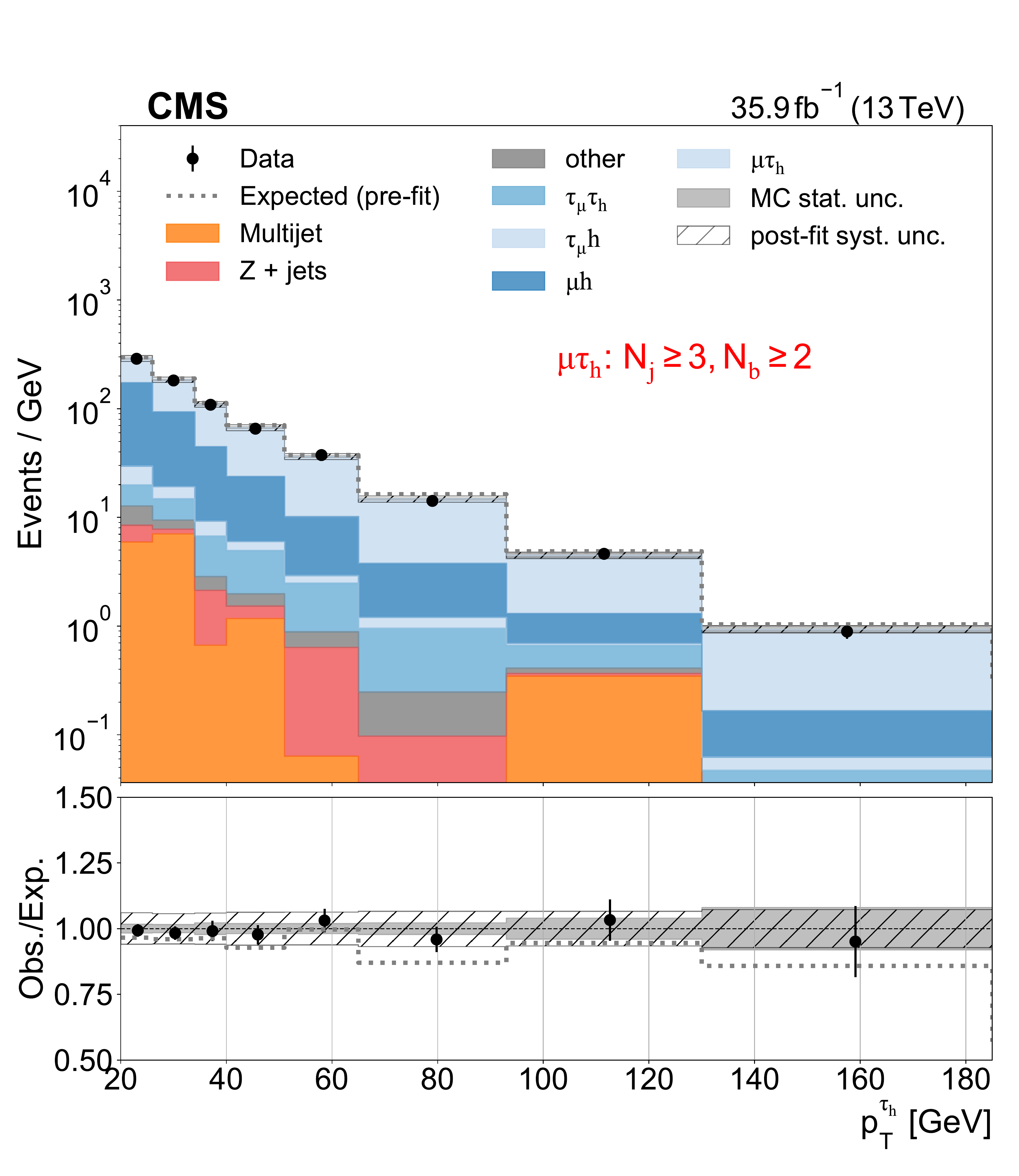}
    
    \caption{Distributions of \tauh \pt used as inputs for the binned likelihood fits for the $\PGm\PGt$ categories. The different panels list the varying selections on the number of jets ($N_{\jet}$) and of \PQb-tagged jets ($N_{\PQb}$) required in each case. The lower subpanels show the ratio of data over pre-fit expectations, with the gray histograms (hatched area) indicating MC statistical (post-fit systematic) uncertainties. Vertical bars on the data markers indicate statistical uncertainties.
    \label{fig:fits_templates_mutau}}
\end{figure*}

\begin{figure*}[htbp!]
    \centering
    \includegraphics[width=0.49\textwidth]{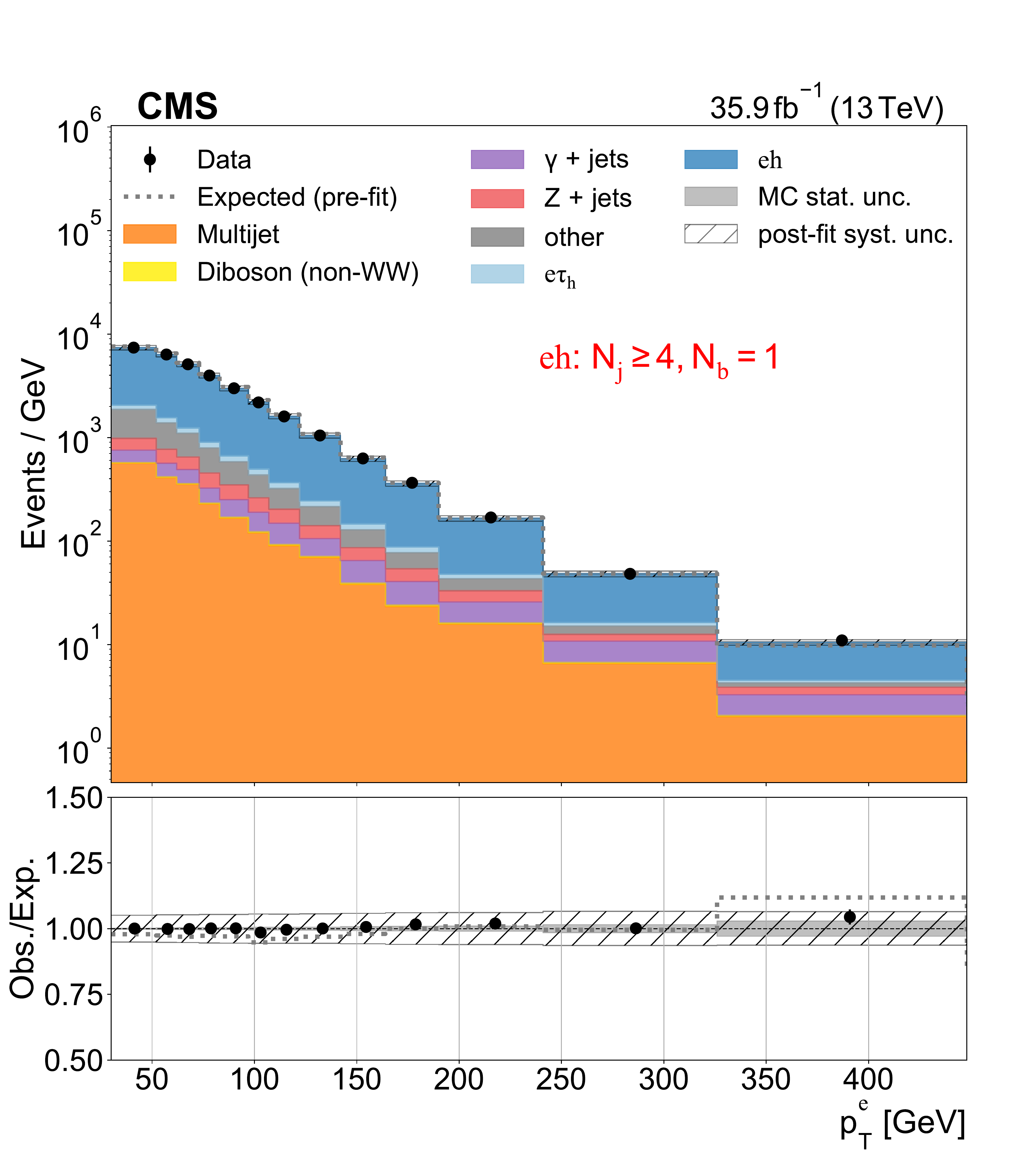}
    \includegraphics[width=0.49\textwidth]{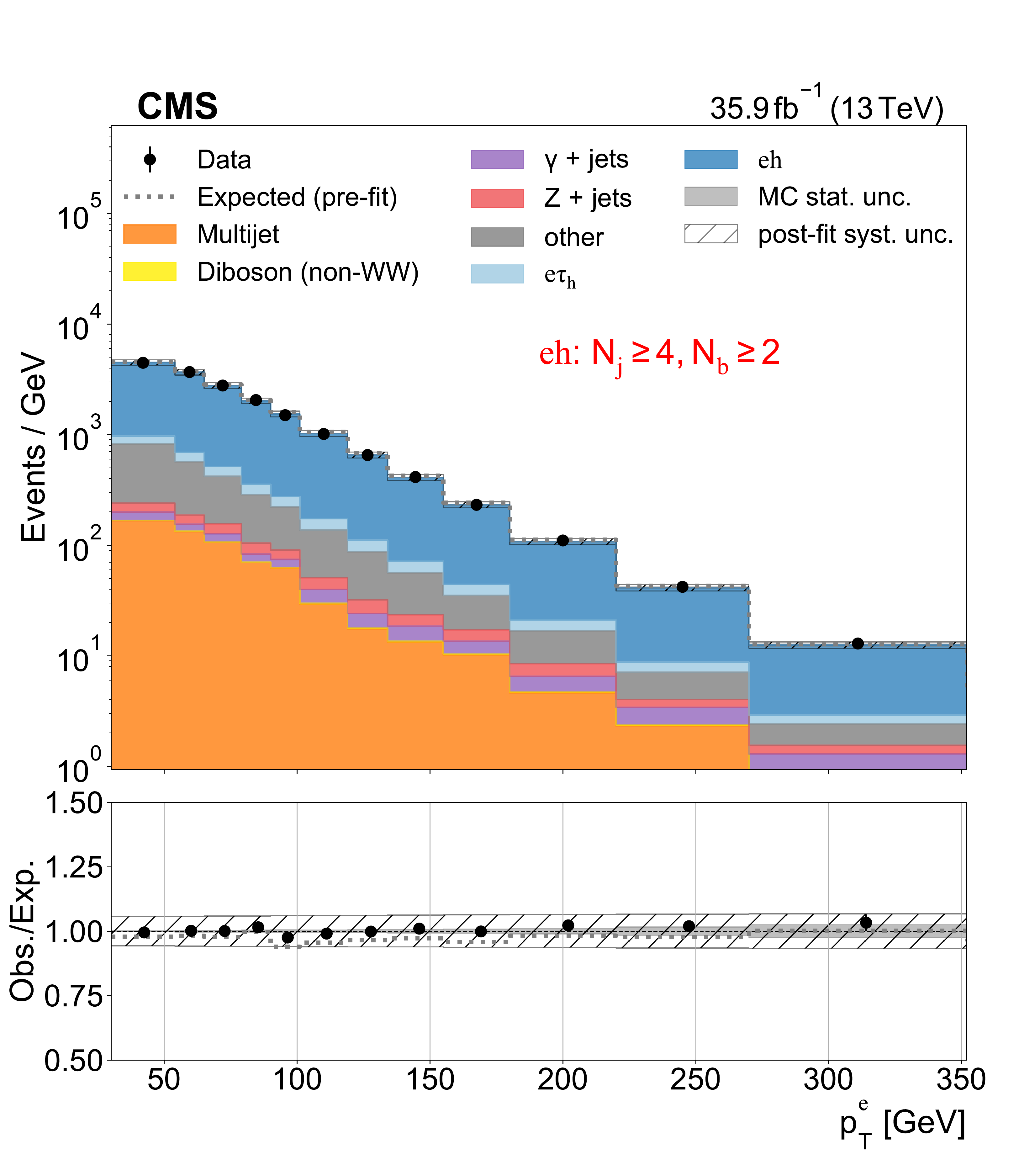}

    \includegraphics[width=0.49\textwidth]{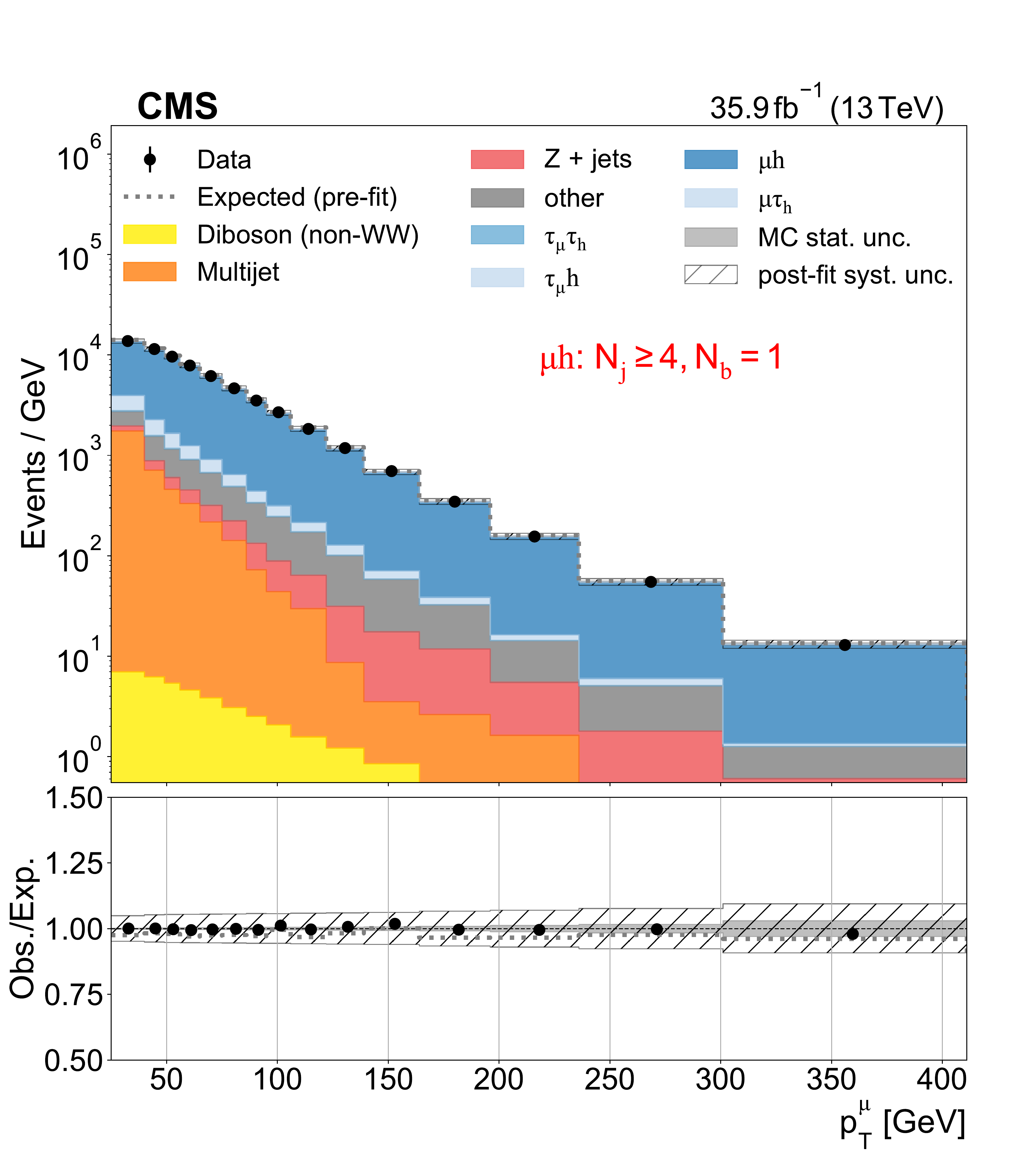}
    \includegraphics[width=0.49\textwidth]{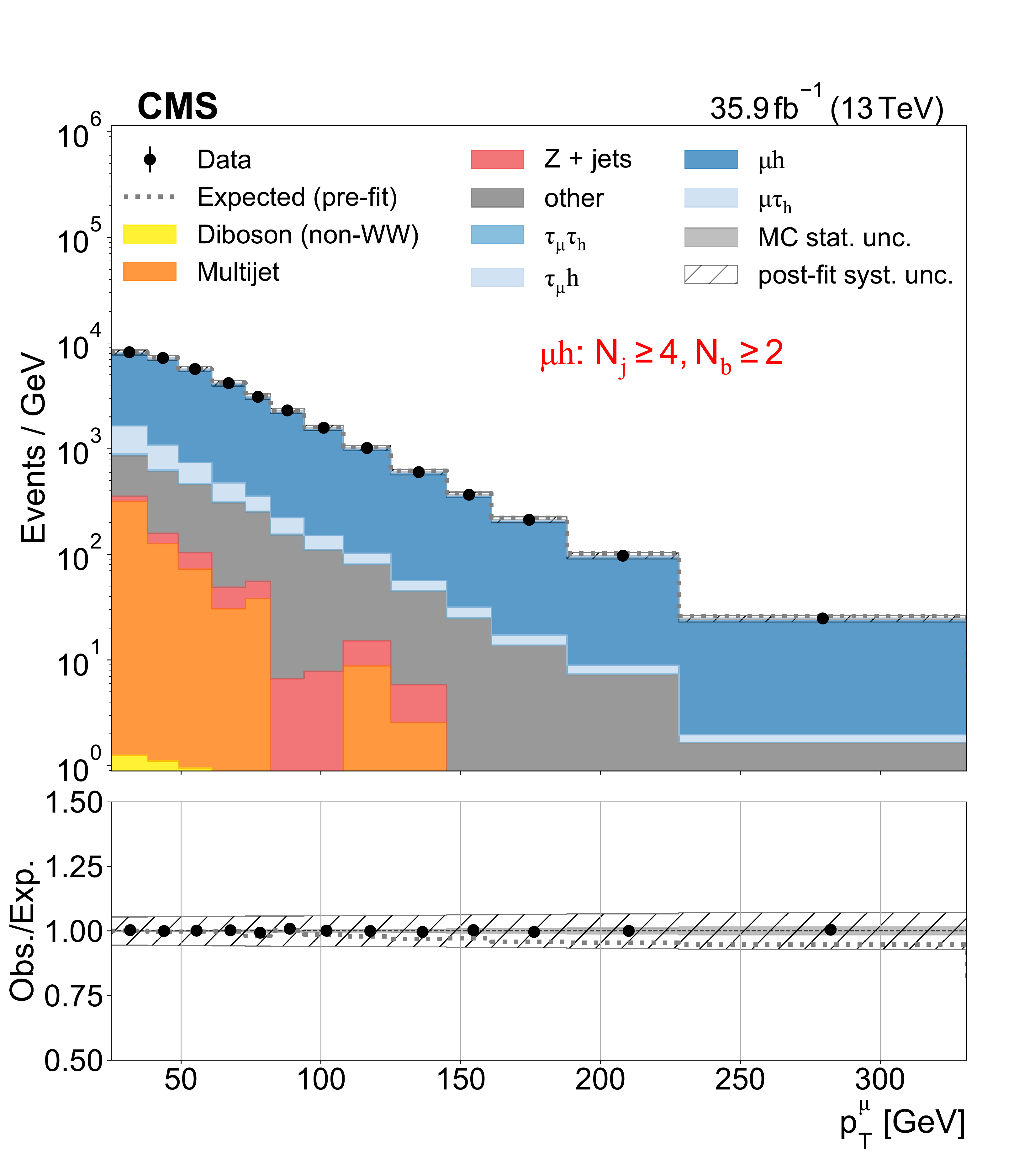}
    \caption{Distributions of electron or muon \pt used as inputs for the binned likelihood fits for the $\Pe \had$ (upper) and $\PGm \had$ (lower) categories, respectively, with the requirement of one (left) or more than one (right) \PQb-tagged jets. The lower subpanels show the ratio of data over pre-fit expectations, with the gray histograms (hatched area) indicating MC statistical (post-fit systematic) uncertainties. Vertical bars on the data markers indicate statistical uncertainties.
    \label{fig:fits_templates_ljet}}
\end{figure*}

The branching fractions (both for the \PW and \PGt decays) are estimated by minimizing Eq.~(\ref{eq:nll_full}) with respect to all parameters over all categories simultaneously.  Because the values of the \PW and \PGt branching fractions are present in the simulation and therefore propagated into the efficiencies, the parameterization of the branching fractions in the likelihood model uses the ratio of fitted branching fractions to their nominal values~\cite{PDG2020}.  Also, because the \PGt branching fractions are known to very high precision and are therefore tightly constrained \textit{a priori}, the fit is insensitive to their values.  The distributions for all considered event categories are shown in Figs.~\ref{fig:fits_templates_ll}--\ref{fig:fits_templates_ljet}. The blueish histograms indicate the simulated contributions expected from signal processes, whereas the red, orange, and yellow ones correspond to different backgrounds. By adding extra requirements on the number of $\PQb$-tagged jets, as can be seen by scanning from left to right, and upper to lower, the panels of each figure, the data distributions are correspondingly more enriched in signal events characterized by increasing production of jets and \PQb jets. In total, there are 30 categories defined by the number and type of reconstructed leptons, the number of jets, and the number of \PQb-tagged jets.  

To cross-check the results derived from the MLE approach, a separate count-based analysis was conducted in parallel. This count-based method did not make use of kinematic information, and included only a subset of event categories that had a high concentration of \ttbar events. For categories that use the same trigger, ratios of the channel yields are constructed that are then analytically solved for the three leptonic branching fractions from a set of quadratic equations. The resulting branching fraction estimates are consistent in both approaches. However, the precision of the count-based method is significantly limited by the \tauh identification systematic uncertainty, and ultimately is less sensitive than the default MLE approach. 
 
\section{Systematic uncertainties}
\label{sec:systematics}

Systematic uncertainties in the MLE fit are accounted for through NPs, denoted with the $\theta$ symbol in Eqs.~(\ref{eq:templates}) and (\ref{eq:nll_full}). The propagation of each individual source of uncertainty is described next.

The uncertainty of the measured value of the CMS integrated luminosity is estimated to be 2.5\%~\cite{CMS-LUM-17-003}. This uncertainty affects the overall normalization of all channels and all simulated processes in a fully correlated manner.

Each simulated event is weighted by a scale factor to account for differences in the pileup spectrum between data and simulation. The uncertainty in the event weights is mainly due to the uncertainty in the total inelastic \pp cross section at 13\TeV~\cite{Sirunyan:2018nqx}, taken as $\sigma_\text{inel} = 69.2 \pm 3.2\unit{mb}$. The effect of this uncertainty is propagated through the analysis by calculating the distribution of pileup in data when varying the $\sigma_\text{inel}$ value up and down by one standard deviation. 

The uncertainties associated with the normalization of the simulated processes with the largest overall contribution to the signal region (\ttbar, Drell--Yan, \WW, and \Wjets) are accounted for by varying the renormalization and factorization scales by a factor of two up and down with respect to their nominal values, and generating the corresponding morphing templates. The NPs are assigned independently for different jet multiplicities such that they are uncorrelated before fitting. The remaining processes (\tW~\cite{CMS:2018amb}, \gjets~\cite{ATLAS:2017xqp}, and non-\WW diboson production~\cite{CMS:2019efc,CMS:2020gtj}) are assigned a single NP each, with a $10\%$ uncertainty in their overall normalization.

The uncertainty in the QCD multijet background estimate from data is included by assigning a channel-dependent ($\Pe\PGm$, $\Pe\tauh$, $\PGm\tauh$, $\Pe\had$, and $\PGm\had$) NP. For the $\Pe\tauh$ and $\PGm\tauh$ channels, the uncertainty is estimated based on comparing the transfer factors between same-sign and opposite-sign events in a region where the light lepton is either isolated or not. For the $\Pe\had$ and $\PGm\had$ categories, the normalization is allowed to vary freely, and consequently is constrained by the data.  In all channels, an NP is assigned for each jet and \PQb tag multiplicity category.

The uncertainties in the efficiency associated with the reconstruction, triggering, identification, and isolation of electrons and muons are accounted for using \pt-dependent NPs that include the statistical as well as the systematic uncertainties from the ``tag-and-probe" procedure~\cite{Khachatryan:2010xn} used to calculate the scale factors.  Additional uncertainty in the trigger efficiency is included for events with electrons in the endcap sections of the detector due to a radiation-induced shift in the ECAL timing in the 2016 data-taking period (referred to as prefiring). To account for the electron and muon energy scales, the lepton \pt that is included in the fitted distribution is varied up and down by one standard deviation and the effect is propagated to the morphing templates.

The \tauh identification and isolation efficiency is accounted for by \pt-dependent NPs, and a 5\% uncertainty~\cite{Sirunyan:2018pgf} is used as a constraint to each bin. The $\mathrm{jet}\to\tauh$ misidentification rate scale factors and uncertainties are derived based on a dilepton plus \tauh candidate control region. An NP is assigned to each \pt bin used to determine the scale factor, and an overall normalization NP is assigned to account for any difference in rate between light- and heavy-quark jets. The case where an electron is misreconstructed as a \tauh candidate, is accounted for by a single normalization NP. The \tauh energy scale is corrected, and an uncertainty of 1.2\% is assigned to it.

The systematic uncertainties associated with the jet energy scale and resolution impact the analysis by modifying the acceptance of events in the various jet multiplicity categories. Their associated uncertainty is derived by varying the jet \pt up and down by one standard deviation for each source of uncertainty associated with the jet energy scale, and assessing the resulting effect on the jet and \PQb-tagged jet multiplicities. The jet energy scale is varied based on a number of contributing uncertainty sources~\cite{Khachatryan:2016kdb}, and incorporated via several shape NPs. The jet energy scale resolution, on the other hand, is treated as a single source of uncertainty based on the associated correction factor.

The \PQb tagging modeling in simulation is corrected with scale factors to better describe the data. The uncertainty in the correction is assessed based on up and down variations of the \PQb tagging and mistag scale factors determined in the multijet enriched control region. The \PQb tagging uncertainties are factorized in the calculation of the scale factors based on their various underlying sources considered. The variation is propagated into the final result through the inclusion of shape NPs for both \PQb tagging and mistag uncertainties.

The uncertainties in the cross sections associated with the PDFs used in the simulation is assessed based on the distribution of weights derived from the 100 NNPDF3.0 replicas. The impact of uncertainty in the value of \alpS is included by considering the effect of its variation within $\alpSmW = 0.1202 \pm 0.0010$~\cite{PDG2020} on both the cross section for each process and, in the case of \ttbar, on the parton showering model via the initial- and final-state radiation (ISR and FSR). The matching of the matrix element calculation to the parton shower is regulated by the $hdamp = 1.38^{+0.93}_{-0.51}$~\cite{Sirunyan:2019dfx} parameter at the generator level. This parameter is varied from its nominal value in dedicated \ttbar MC samples to estimate its effects on the normalization and on the fitted distributions. Uncertainties related to the modeling of the underlying event are derived from dedicated \PYTHIA CUETP8M2T4 tune analyses~\cite{CMS-PAS-TOP-16-021}.  Several differential measurements of the \ttbar cross section have observed a \pt distribution of the top quark that is softer than predicted by the \POWHEG simulation~\cite{CMS:2018jcg, Sirunyan:2017mzl, Khachatryan:2016mnb}. To account for any top \pt distribution mismodeling, an uncertainty is assigned based on reweighting simulation to data and deriving a one-sided prior distribution from the difference with respect to the nominal simulation.  The \pt spectrum of \WW events generated with \POWHEG is reweighted to match the analytical prediction obtained using \pt-resummation at next-to-next-to-leading logarithmic accuracy~\cite{Grazzini:2015wpa}, and the associated uncertainties are assessed by varying the resummation, factorization, and renormalization scales in the analytical calculation~\cite{Meade:2014fca}.

\begin{table*}[htpb!]
    \centering
    \setlength{\tabcolsep}{0.5em}
    \topcaption{Summary of the impacts of each source of uncertainty (quoted as a percent of the total systematic uncertainty) for each \PW branching fraction. Whenever multiple NPs impact a common source of systematic uncertainty, each component is varied independently and the range of impacts is given.}
    \label{tab:systematics_impacts}
    \begin{scotch}{lcccc}
                                                      & $\PW\to \Pe\PAGne$ & $\PW\to \PGm\PAGnGm$ & $\PW\to \PGt\PAGnGt$ & $\PW\to \qqbar'$ \\
        \hline
        Pileup                                        & 20       & 6        & 11       & 14       \\
        Luminosity                                    & 5        & 14       & 5        & 7        \\
        JES/JER                                       & 3--17    & 5--21    & 4--11    & 4--21    \\
        $\PQb$ tagging                                & $<$1--19 & $<$1--25 & $<$1--5  & $<$1--17 \\
        $\cPqt\PW$ normalization                      & 35       & 43       & 27       & 46       \\
        $\PW\PW$ normalization                        & 8        & 9        & 5        & 9        \\
        $\PW\PW$ \pt                                  & 1--2     & 1--2     & $<$1--5  & $<$1--4  \\
        $\PW+$\,jets normalization                    & $<$1--6  & $<$1--7  & $<$1--13 & $<$1--10 \\
        $\Pgg+$\,jets normalization                   & 1        & 2        & 5        & 4        \\
        $\PW\PZ,\PZ\PZ$ normalization                 & $<$1     & 1        & $<$1     & $<$1     \\[\cmsTabSkip]
  
        \multicolumn{5}{l}{\ttbar production:}       \\
        \hspace{3mm} QCD scale                        & 32       & 47       & 25       & 45       \\
        \hspace{3mm} top quark \pt                          & 16       & 24       & 7        & 18       \\
        \hspace{3mm} ISR                              & 10       & 16       & 37       & 37       \\
        \hspace{3mm} FSR                              & 3        & 4        & 9        & 5        \\
        \hspace{3mm} PDF                              & 4        & 5        & 3        & 4        \\
        \hspace{3mm} \alpS                            & 5        & 5        & 3        & 6        \\
        \hspace{3mm} \PYTHIA 8 UE tune                & 1        & 5        & 7        & 7        \\
        \hspace{3mm} $hdamp$ parameter                & 3        & 3        & 2        & 4        \\[\cmsTabSkip]
  
        \multicolumn{5}{l}{Drell--Yan background:}   \\
        \hspace{3mm} QCD scale                        & 2--24    & 10--27   & 5--20    & 8--30    \\
        \hspace{3mm} PDF                              & 3        & 5        & 2        & 4        \\[\cmsTabSkip]
  
        \multicolumn{5}{l}{QCD multijet background:} \\
        \hspace{3mm} $\Pe\PGm$                        & 5        & 12       & 12       & 6        \\
        \hspace{3mm} $\Pe\had$                  & 3--4     & 11--17   & 6--7     & 6--10    \\
        \hspace{3mm} $\PGm\had$                 & 10--11   & 10--13   & 5--13    & 2--3     \\
        \hspace{3mm} $\Pe\tauh$                       & $<$1--5  & $<$1--8  & $<$1--9  & $<$1--7  \\
        \hspace{3mm} $\PGm\tauh$                      & $<$1--12 & $<$1--10 & $<$1--9  & $<$1--10 \\[\cmsTabSkip]
  
        \multicolumn{5}{l}{\Pe measurement:}         \\
        \hspace{3mm} Reconstruction efficiency        & 50       & 13       & 3        & 15       \\
        \hspace{3mm} Identification efficiency        & $<$1--14 & 1--8     & $<$1--10 & $<$1--5  \\
        \hspace{3mm} Trigger (prefiring)              & 29       & 2        & 1        & 9        \\
        \hspace{3mm} Trigger                          & $<$1--27 & $<$1--4  & $<$1--13 & $<$1--9  \\
        \hspace{3mm} Energy scale                     & 7        & 6        & $<$1     & 4        \\[\cmsTabSkip]
  
        \multicolumn{5}{l}{\PGm measurement:}        \\
        \hspace{3mm} Reconstruction efficiency        & $<$1--2  & $<$1--5  & $<$1--6  & $<$1--6  \\
        \hspace{3mm} Trigger                          & 8        & 26       & 3        & 7        \\
        \hspace{3mm} Energy scale                     & 1        & $<$1     & 3        & 2        \\[\cmsTabSkip]
  
        \multicolumn{5}{l}{\tauh measurement:}        \\
        \hspace{3mm} Reconstruction efficiency        & 2--14    & 7--17    & 21--46   & 14--24   \\
        \hspace{3mm} Energy scale                     & 9        & 5        & 14       & 6        \\
        \hspace{3mm} Jet misidentification            & 1--14    & $<$1--10 & 1--24    & $<$1--10 \\
        \hspace{3mm} \Pe misidentification            & $<$1     & $<$1     & 2        & 1        \\
        \hspace{3mm} $\PGt\to\Pe, \PGm, \had$            & $<$1     & $<$1     & $<$1--2  & $<$1--1  \\
    \end{scotch}
\end{table*}

The impacts on the measured values of the branching fractions from each uncertainty source are estimated by individually varying each NP both up and down by one standard deviation based on their post-fit uncertainties, carrying out the fit with the NP under consideration fixed to the varied value, and then evaluating the corresponding change in each of the branching fractions with respect to their central MLE values. These impacts are summarized in Table~\ref{tab:systematics_impacts} where the values reported indicate the magnitude of the change in each measured branching fraction normalized by the total uncertainty of each branching fraction. A range of values is quoted in cases where multiple NPs are assigned to a systematic uncertainty source, and the scale of the impact changes depending on the NP being varied. The quoted impacts do not need to add up to 100\% of the branching fraction uncertainty given the correlations among them (the individual uncertainties represented by the impacts would need to be summed in quadrature to equal the total variance). The most important sources of uncertainties are the \ttbar, $\PQt\PW$, and Drell--Yan normalizations, as well as the top-quark ISR and \pt modeling---common to all \PW branching fraction extractions---and the electron reconstruction efficiency, the \PGm triggering, and the \tauh reconstruction efficiency, for the electron, muon, and \PGt branching fraction determinations, respectively.
        
\section{Results}

\begin{figure}[htbp!]
    \centering

        \includegraphics[width=\cmsFigWidth]{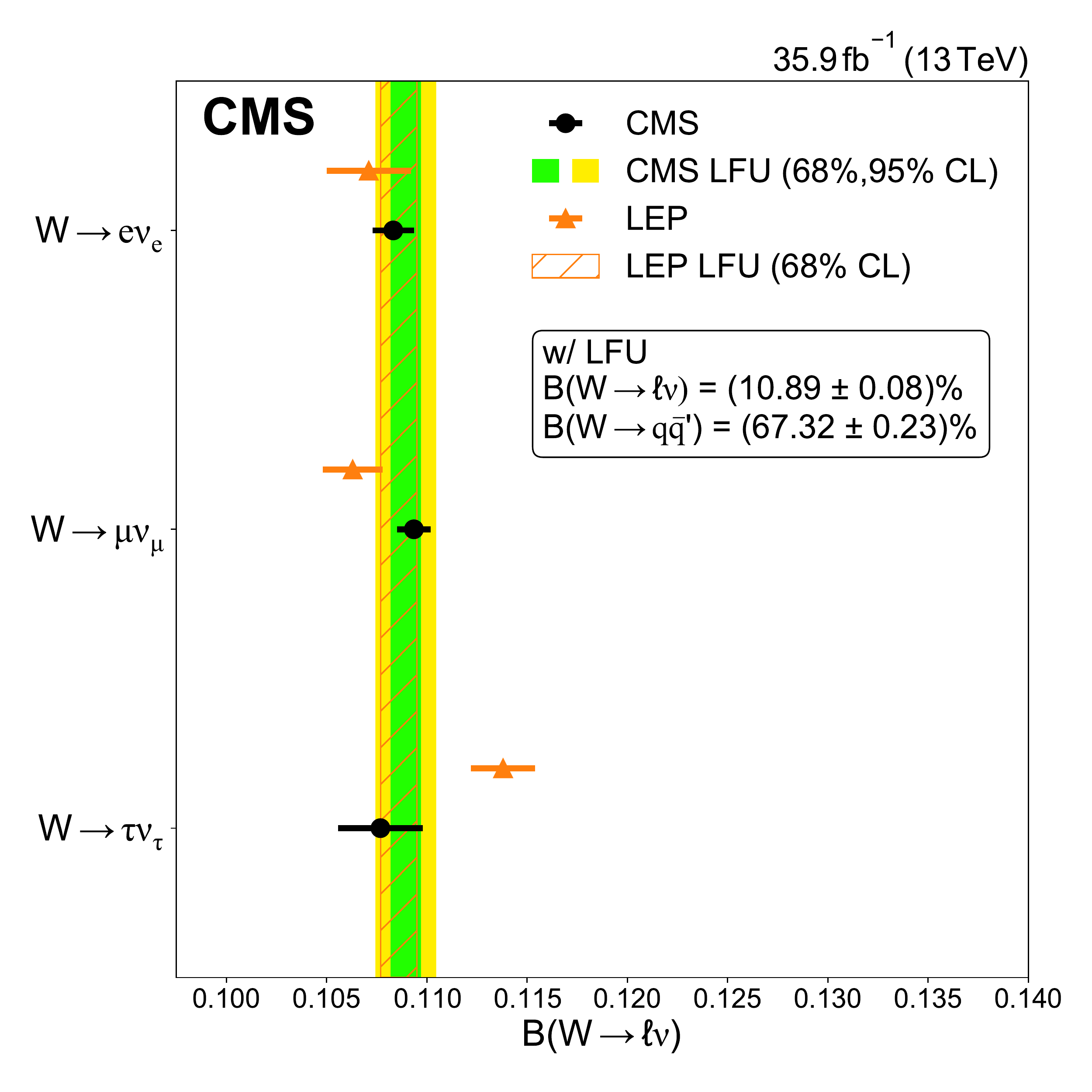}

    \caption{Summary of the measured values of the \PW leptonic branching fractions compared with the corresponding LEP results~\cite{Schael:2013ita,PDG2020}. The vertical green-yellow band shows the extracted \PW leptonic branching fraction assuming LFU (the hatched band shows the corresponding LEP result).  The horizontal error bars on the data points indicate their total uncertainty.
    \label{fig:wbr_result_1D}}
\end{figure}

The values of the branching fractions obtained as described in the previous sections are shown in Table~\ref{tab:results} for the scenario where each leptonic branching fraction in the MLE fit can vary independently, and where they are all fixed to the same value according to LFU. The results are also plotted in Fig.~\ref{fig:wbr_result_1D}, together with the corresponding values determined from a combination of the LEP measurements~\cite{Schael:2013ita,PDG2020}. The green (yellow) bands in this plot, and in all figures hereafter, indicate the 68\% (95\%) confidence level (\CL) results for the extracted branching fractions. Whereas the systematic uncertainties of the CMS and LEP measurements are similar, the extractions reported here are 3--10 times more precise statistically than those from LEP. The final electron and muon branching fractions are thereby measured about 1.5 times more precisely than at LEP, whereas the \PGt lepton extractions have similar total uncertainty but mostly of systematic (statistical) origin in the CMS (LEP) case. Under the LFU assumption, an average leptonic decay branching fraction of $\brWtolnu = (10.89 \pm 0.01 \pm 0.08)\%$ is derived, where the first and second uncertainties correspond to the statistical and systematic sources, respectively. This result is consistent with, but much more statistically precise than, the value of $(10.86 \pm 0.06 \pm 0.09)\%$ obtained from the LEP data. The inclusive hadronic \PW boson decay branching fraction, $\brWtoqq = (67.32 \pm 0.02 \pm 0.23)\%$, is obtained by imposing the constraint $\brWtoqq= 1 - 3\brWtolnu$ in the likelihood. The resulting uncertainty is approximately 15\% smaller than at LEP.

\begin{figure*}[htbp!]
    \centering
    \includegraphics[width=1.0\textwidth]{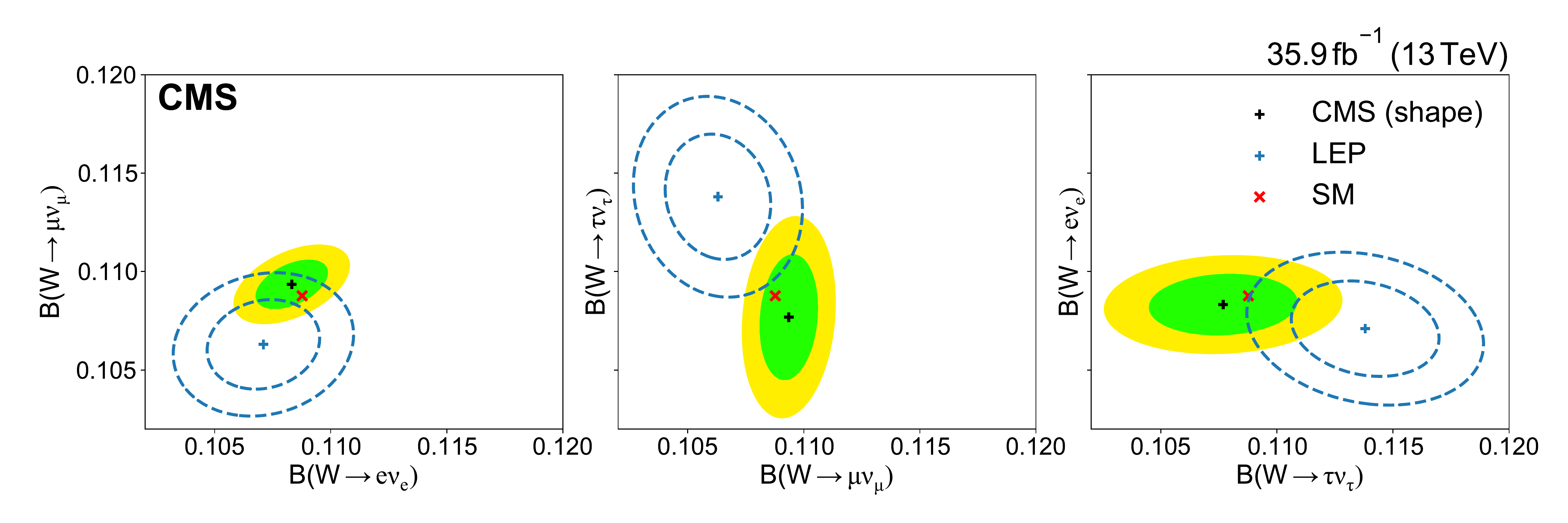}
    \caption{Two-dimensional distributions of pairs of \PW leptonic branching fractions derived here compared with the corresponding LEP results~\cite{Schael:2013ita,PDG2020} and to the SM expectation. The green (darker) and yellow (lighter) bands (dashed lines for the LEP results) correspond to the 68\% and 95\% \CL, respectively, for the resulting two-dimensional Gaussian distribution.
    \label{fig:contours_2D}}
\end{figure*}

\begin{figure}[htbp!]
    \centering
        \includegraphics[width=\cmsFigWidth]{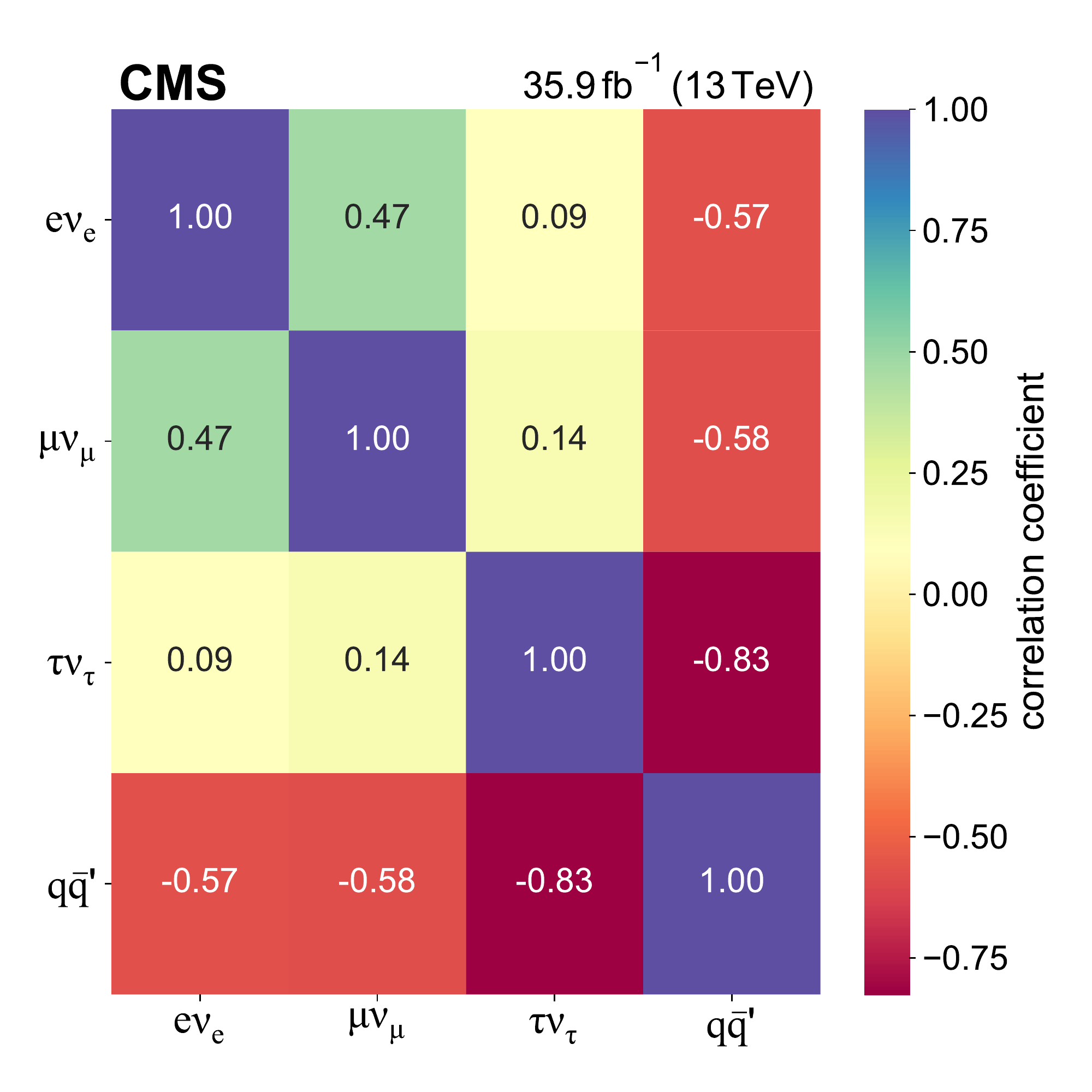}
    \caption{Correlation matrix between the four \PW boson decay branching fraction components extracted in this work.
    \label{fig:correlation_matrix_POI}}
\end{figure}

The individually extracted branching fractions are strongly correlated because of the composition of the selected data samples, and because of the constraint that the sum of leptonic and hadronic branching fractions is unity. To demonstrate the pairwise correlations between leptonic branching fractions, two-dimensional contours are shown in Fig.~\ref{fig:contours_2D}. For each pair shown in the panels, the third branching fraction that is not plotted has been integrated out. Additionally, the correlation matrix associated to the branching fraction measurements is shown in Fig.~\ref{fig:correlation_matrix_POI}. The \brWtoqq and \brWtotaunu branching fractions have the largest (anti)correlation ($-0.83$), whereas \brWtoenu and \brWtotaunu appear to be the least correlated quantities ($0.09$ correlation factor).

\begin{table*}[htbp!]
    \centering
    \topcaption{Values of the \PW boson decay branching fractions measured here compared with the corresponding LEP measurements~\cite{Schael:2013ita,PDG2020}. The lower rows list the average leptonic and inclusive hadronic \PW branching fractions derived assuming LFU. The first and second uncertainties quoted for each branching fraction correspond to statistical and systematic sources, respectively.
    \label{tab:results}}
    \begin{scotch}{l ccc}
                 & CMS                            & LEP                            \\
    \hline
    \brWtoenu    & $(10.83 \pm 0.01 \pm 0.10) \%$ & $(10.71 \pm 0.14 \pm 0.07)$ \% \\
    \brWtomunu   & $(10.94 \pm 0.01 \pm 0.08) \%$ & $(10.63 \pm 0.13 \pm 0.07)$ \% \\
    \brWtotaunu  & $(10.77 \pm 0.05 \pm 0.21) \%$ & $(11.38 \pm 0.17 \pm 0.11)$ \% \\
    \brWtoqq     & $(67.46 \pm 0.04 \pm 0.28) \%$ & \NA                             \\[\cmsTabSkip]
 
    Assuming LFU &                                &                                 &  \\[\cmsTabSkip]
 
    \brWtolnu    & $(10.89 \pm 0.01 \pm 0.08)\%$  & $(10.86 \pm 0.06 \pm 0.09)\%$  \\
    \brWtoqq     & $(67.32 \pm 0.02 \pm 0.23)\%$  & $(67.41 \pm 0.18 \pm 0.20)\%$  \\
    \end{scotch}
\end{table*}

\begin{figure}[htbp!]
    \centering
        \includegraphics[width=\cmsFigWidth]{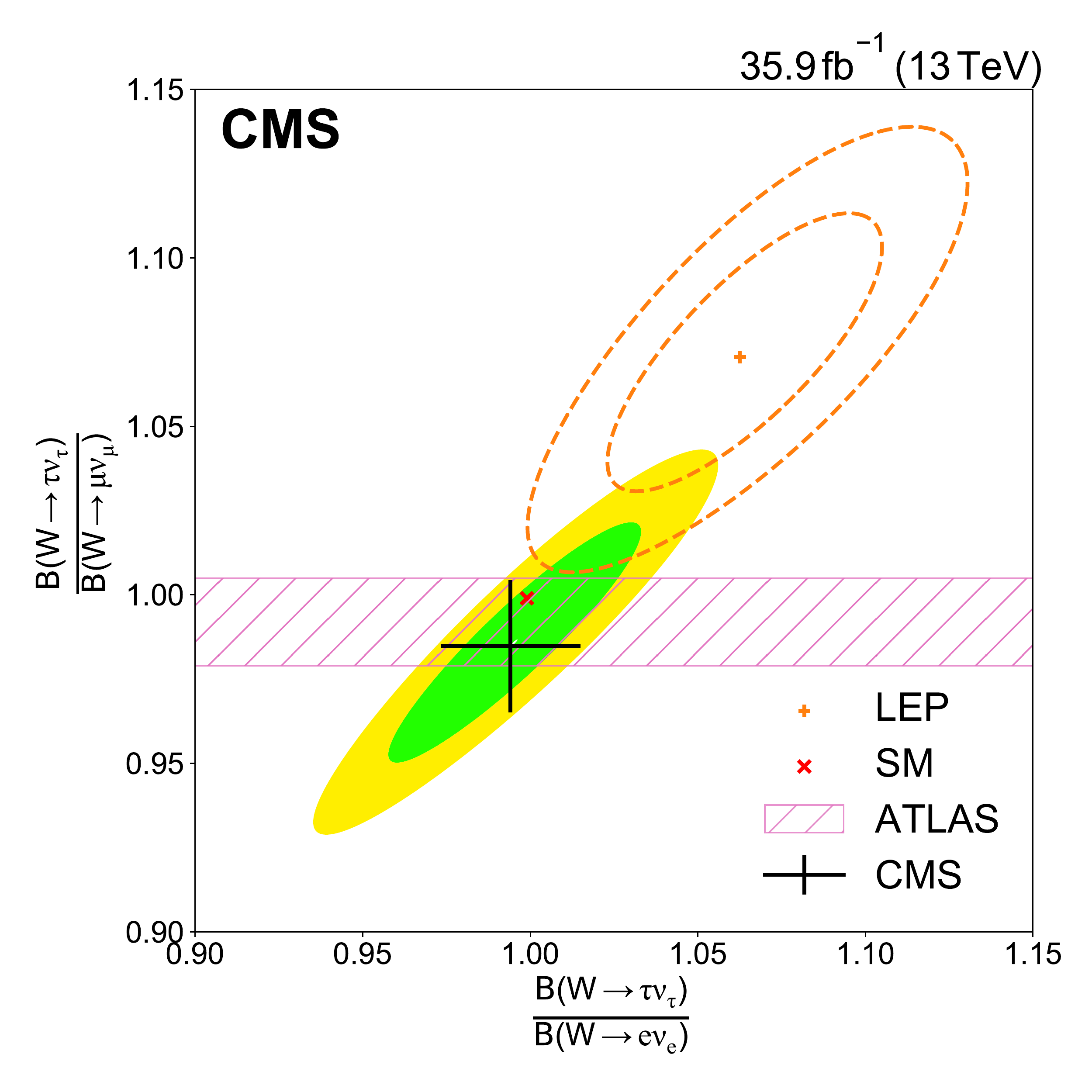}
    \caption{Two-dimensional distribution of the ratio \Rtaue versus \Rtaumu, compared with the corresponding LEP~\cite{Schael:2013ita,PDG2020} and ATLAS~\cite{Aad:2020ayz} results and with the SM expectation. The green and yellow bands (dashed lines for the LEP results) correspond to the 68\% and 95\% \CL, respectively, for the resulting two-dimensional Gaussian distribution. The corresponding 68\% \CL one-dimensional projections (black error bars) are also overlaid for a better visual comparison with the ATLAS \Rtaumu result. 
    \label{fig:ratios_2D}}
\end{figure}

Having measured the branching fractions, it is of interest to calculate the ratios among them with their associated probability distribution functions (pdfs) to compare those with similar results from other experiments where only such ratios have been measured. To transform the likelihood of the branching fractions, ${B}_{\ell} \equiv \brWtolnu$, to the likelihood of their ratios, $R_{\ell'/\ell}$, the following integral transformation is evaluated~\cite{10.2307/2334671} 
\begin{linenomath} 
\begin{equation}
    f(R_{\ell'/\ell}) = \int_{-\infty}^{\infty} \abs{\mathcal{B}_{\ell}} g(R_{\ell'/\ell} \mathcal{B}_{\ell}, \mathcal{B}_{\ell}) \,\rd B_{\ell},
\end{equation}
\end{linenomath}
where the pdf of the branching fractions $g(\mathcal{B}_{\ell'}, \mathcal{B}_\ell)$ is a bivariate normal distribution with parameters determined from the likelihood fit. It is also possible to carry out the transformation above in the two-dimensional case, so that ratios of \PGt lepton over muon and electron decays can be compared between each other as well as with the SM expectation, as shown in Fig.~\ref{fig:ratios_2D}. Table~\ref{tab:ratios} lists the ratios obtained as described above, compared with those measured at LEP, LHC, and Tevatron.
The ATLAS \Rtaumu extraction~\cite{Aad:2020ayz} has a smaller uncertainty than that of CMS because it benefits, in part, from a four times larger \pp data sample analyzed. Within the current uncertainties, all CMS ratios are consistent with the LFU hypothesis given by $R_{\ell/\ell'}\approx 1$.

\begin{table*}[htbp!]
    \centering
    \topcaption{Ratios of different leptonic branching fractions, $R_{\PGm/\Pe} = \mathcal{B}(\PW\to \PGm\PAGnGm)/\mathcal{B}(\PW\to \Pe\PAGne)$,  $R_{\PGt/\Pe} = \mathcal{B}(\PW\to \PGt\PAGnGt)/\mathcal{B}(\PW\to \Pe\PAGne)$, and  $R_{\PGt/\PGm} = \mathcal{B}(\PW\to \PGt\PAGnGt)/\mathcal{B}(\PW\to \PGm\PAGnGm)$, measured here compared with the values obtained by other LEP~\cite{Schael:2013ita}, LHC~\cite{ATLAS:2016nqi,LHCb:2016zpq,Aad:2020ayz}, and Tevatron~\cite{D0:1995gzy,CDF:2005bdv} experiments.
    \label{tab:ratios}}
    \cmsTable{
        \begin{scotch}{c cccccc}
                        & CMS               & LEP               & ATLAS             & LHCb              & CDF               & D0 \\
        \hline
        $R_{\PGm/\Pe}$  & $1.009 \pm 0.009$ & $0.993 \pm 0.019$ & $1.003 \pm 0.010$ & $0.980 \pm 0.012$ & $0.991 \pm 0.012$ & $0.886 \pm 0.121$ \\
        $R_{\PGt/\Pe}$  & $0.994 \pm 0.021$ & $1.063 \pm 0.027$ & \NA                & \NA                & \NA                & \NA \\
        $R_{\PGt/\PGm}$ & $0.985 \pm 0.020$ & $1.070 \pm 0.026$ & $0.992 \pm 0.013$ & \NA                & \NA                & \NA \\
        $R_{\PGt/\ell}$ & $1.002 \pm 0.019$ & $1.066 \pm 0.025$ & \NA                & \NA                & \NA                & \NA \\
        \end{scotch}
}
\end{table*}

From the determined values of the average leptonic and inclusive hadronic \PW branching fractions, and following Eq.~(\ref{eq:W_had}), other interesting SM quantities can be derived such as the QCD coupling constant at the \PW boson mass scale, $\alpS(m^{2}_\PW)$, or the $\abs{V_{{\PQc\PQs}}}$ CKM element. One can similarly check the unitarity of the first two rows of the CKM matrix, given by the squared sum in the prefactor of Eq.~(\ref{eq:W_had}). To extract those SM parameters, one compares the measured ratio of hadronic-to-leptonic branching fractions to the corresponding theoretical expression, parameterized at next-to-next-to-next-to-leading-order QCD plus LO EW and mixed EW+QCD accuracy~\cite{dEnterria:2020cpv}, leaving either $\alpS(m^{2}_\PW)$ or the (sum of) CKM matrix element(s) free, using the following expression:
\begin{linenomath}
\label{eq:br_to_ckm}
\ifthenelse{\boolean{cms@external}}
{
	\begin{multline}
		\frac{\mathcal{B}(\PW\to\qqbar')}{1 - \mathcal{B}(\PW\to\qqbar')} = \sum_{\substack{{i} = (\PQu,\PQc), \\ j =(\PQd, \PQs, \PQb)}}\abs{V_{{ij}}}^{2} \\ \times \Bigl[1 + \sum_{i=1}^4 c_i\left(\frac{\alpS}{\pi}\right)^i + c_\mathrm{EW}(\alpha) + c_{\text{mix}}(\alpha \alpS) \Bigr],
	\end{multline}
}{
    \begin{equation}
        \frac{\mathcal{B}(\PW\to\qqbar')}{1 - \mathcal{B}(\PW\to\qqbar')} = \sum_{\substack{{i} = (\PQu,\PQc), \\ j =(\PQd, \PQs, \PQb)}}\abs{V_{{ij}}}^{2} \Bigl[1 + \sum_{i=1}^4 c_i\left(\frac{\alpS}{\pi}\right)^i + c_\mathrm{EW}(\alpha) + c_{\text{mix}}(\alpha \alpS) \Bigr],
    \end{equation}
}
\end{linenomath}
where the numerical value of the ratio derived from the experimental result presented here is $2.060\pm 0.021$. The theoretical uncertainties of Eq.~(\ref{eq:br_to_ckm}), from parametric dependencies and missing higher-order corrections~\cite{dEnterria:2016rbf,dEnterria:2020cpv}, are much smaller than the experimental uncertainty of this ratio. If CKM unitarity is imposed, then the sum in Eq.~(\ref{eq:br_to_ckm}) is $\sum_{ij}\abs{V_{{ij}}}^{2} = 2$ and a value of $\alpS(m^{2}_\PW) = 0.095 \pm 0.033$ can be inferred. This value is much less precise than the current world-average QCD coupling constant, which amounts to $\alpS(m^{2}_\PW) = 0.1202 \pm 0.0010$ at the \PW boson mass scale~\cite{PDG2020}, but confirms the usefulness of \PW boson hadronic decays to extract this fundamental parameter at future \EE colliders where the \PW boson branching fractions can be measured much more precisely~\cite{Abada:2019zxq}. If, instead, the current world average of $\alpS(m^{2}_\PW)$ is used in Eq.~(\ref{eq:br_to_ckm}), and the sum in Eq.~(\ref{eq:br_to_ckm}) is left free, a value of $\sum_{ij}\abs{V_{{ij}}}^{2} = 1.984 \pm 0.021$ is obtained that provides a precise test of CKM unitarity. Further solving Eq.~(\ref{eq:br_to_ckm}) for $\abs{V_{{\PQc\PQs}}}$, and using the more precisely measured values of the other CKM matrix elements~\cite{PDG2020} in the sum, yields a value of $\abs{V_{{\PQc\PQs}}} = 0.967\pm 0.011$ that is as precise as the value $\abs{V_{{\PQc\PQs}}} = 0.987 \pm 0.011$ directly measured from semileptonic \PD or leptonic \PDs decays, using lattice QCD calculations of the semileptonic \PD form factor or the \PDs decay constant~\cite{PDG2020}. The precision extracting the $\alpS(m^{2}_\PW)$ and $\abs{V_{{\PQc\PQs}}}$ parameters, as well as the CKM unitarity test, is virtually entirely determined by the systematic uncertainty of the average leptonic branching fraction measurement assuming LFU.  A summary of the values calculated here are presented in Table~\ref{tab:alphas_CKM}.  The full tabulated results are provided in HEPData~\cite{hepdata}.

\begin{table}[htb!]
	\centering
	\topcaption{Values of the QCD coupling constant at the \PW mass, the charm-strange CKM mixing element, and the squared sum of the first two rows of the CKM matrix, derived in this work.
	\label{tab:alphas_CKM}}
	\begin{scotch}{ccc}
	\alpSmW           & $\abs{\Vcs}$     & $\sum_{ij}\abs{\Vij}^{2}$ \\ \hline
	$0.095 \pm 0.033$ & $0.967\pm 0.011$ & $1.984 \pm 0.021$
    \end{scotch}
\end{table}

\section{Summary}

A precise measurement of the three leptonic decay branching fractions of the \PW boson has been presented, as well as the average leptonic and inclusive hadronic branching fractions assuming lepton flavor universality (LFU). The analysis is based on a data sample of \pp collisions at a center-of-mass energy of 13\TeV corresponding to an integrated luminosity of 35.9\fbinv recorded by the CMS experiment. Events with one or two \PW bosons produced are collected using single-charged-lepton triggers that require at least one prompt electron or muon with large transverse momentum. The extraction of the \PW boson leptonic branching fractions is performed through a binned maximum likelihood fit of events split into multiple categories defined based on the multiplicity and flavor of reconstructed leptons, the number of jets, and the number of jets identified as originating from the hadronization of \PQb quarks. The measured branching fractions for the decay of the \PW boson into electrons, muons, tau leptons, and hadrons are $(10.83 \pm 0.10)\%$, $(10.94 \pm 0.08)\%$, $(10.77 \pm 0.21)\%$, and $(67.46 \pm 0.28)\%$, respectively. These results are consistent with the LFU hypothesis for the weak interaction, and are more precise than previous measurements based on data collected by the LEP experiments.

Fitting the data assuming LFU provides values of $(10.89 \pm 0.08)\%$ and $(67.32 \pm 0.23)\%$, respectively, for the average leptonic and inclusive hadronic branching fractions of the \PW boson. The comparison of the ratio of hadronic-to-leptonic branching fractions to the theoretical prediction is used to derive other standard model quantities. A value of the strong coupling constant at the \PW boson mass scale of $\alpS(m^{2}_\PW) = 0.095 \pm 0.033$ is obtained which, although not competitive compared with the current world average, confirms the usefulness of the \PW boson decays to constrain this fundamental standard model parameter at future colliders.  Using the world average value of $\alpS(m^{2}_\PW)$, the sum of the square of the elements in the first two rows of the Cabibbo--Kobayashi--Maskawa (CKM) matrix is $\sum_{ij}\abs{V_{{ij}}}^{2} = 1.984 \pm 0.021$, providing a precise check of CKM unitarity. From this sum and using the world-average values of the other relevant CKM matrix elements, a value of $\abs{V_{{\PQc\PQs}}} = 0.967 \pm 0.011$ is determined, which is as precise as the current $\abs{V_{{\PQc\PQs}}} = 0.987 \pm 0.011$ result obtained from direct \PD meson decay data. 

\begin{acknowledgments}
    We congratulate our colleagues in the CERN accelerator departments for the excellent performance of the LHC and thank the technical and administrative staffs at CERN and at other CMS institutes for their contributions to the success of the CMS effort. In addition, we gratefully acknowledge the computing centers and personnel of the Worldwide LHC Computing Grid and other centers for delivering so effectively the computing infrastructure essential to our analyses. Finally, we acknowledge the enduring support for the construction and operation of the LHC, the CMS detector, and the supporting computing infrastructure provided by the following funding agencies: BMBWF and FWF (Austria); FNRS and FWO (Belgium); CNPq, CAPES, FAPERJ, FAPERGS, and FAPESP (Brazil); MES and BNSF (Bulgaria); CERN; CAS, MoST, and NSFC (China); MINCIENCIAS (Colombia); MSES and CSF (Croatia); RIF (Cyprus); SENESCYT (Ecuador); MoER, ERC PUT and ERDF (Estonia); Academy of Finland, MEC, and HIP (Finland); CEA and CNRS/IN2P3 (France); BMBF, DFG, and HGF (Germany); GSRI (Greece); NKFIA (Hungary); DAE and DST (India); IPM (Iran); SFI (Ireland); INFN (Italy); MSIP and NRF (Republic of Korea); MES (Latvia); LAS (Lithuania); MOE and UM (Malaysia); BUAP, CINVESTAV, CONACYT, LNS, SEP, and UASLP-FAI (Mexico); MOS (Montenegro); MBIE (New Zealand); PAEC (Pakistan); MSHE and NSC (Poland); FCT (Portugal); JINR (Dubna); MON, RosAtom, RAS, RFBR, and NRC KI (Russia); MESTD (Serbia); MCIN/AEI and PCTI (Spain); MOSTR (Sri Lanka); Swiss Funding Agencies (Switzerland); MST (Taipei); ThEPCenter, IPST, STAR, and NSTDA (Thailand); TUBITAK and TAEK (Turkey); NASU (Ukraine); STFC (United Kingdom); DOE and NSF (USA).
    
    \hyphenation{Rachada-pisek} Individuals have received support from the Marie-Curie program and the European Research Council and Horizon 2020 Grant, contract Nos.\ 675440, 724704, 752730, 758316, 765710, 824093, 884104, and COST Action CA16108 (European Union); the Leventis Foundation; the Alfred P.\ Sloan Foundation; the Alexander von Humboldt Foundation; the Belgian Federal Science Policy Office; the Fonds pour la Formation \`a la Recherche dans l'Industrie et dans l'Agriculture (FRIA-Belgium); the Agentschap voor Innovatie door Wetenschap en Technologie (IWT-Belgium); the F.R.S.-FNRS and FWO (Belgium) under the ``Excellence of Science -- EOS" -- be.h project n.\ 30820817; the Beijing Municipal Science \& Technology Commission, No. Z191100007219010; the Ministry of Education, Youth and Sports (MEYS) of the Czech Republic; the Deutsche Forschungsgemeinschaft (DFG), under Germany's Excellence Strategy -- EXC 2121 ``Quantum Universe" -- 390833306, and under project number 400140256 - GRK2497; the Lend\"ulet (``Momentum") Program and the J\'anos Bolyai Research Scholarship of the Hungarian Academy of Sciences, the New National Excellence Program \'UNKP, the NKFIA research grants 123842, 123959, 124845, 124850, 125105, 128713, 128786, and 129058 (Hungary); the Council of Science and Industrial Research, India; the Latvian Council of Science; the Ministry of Science and Higher Education and the National Science Center, contracts Opus 2014/15/B/ST2/03998 and 2015/19/B/ST2/02861 (Poland); the Funda\c{c}\~ao para a Ci\^encia e a Tecnologia, grant CEECIND/01334/2018 (Portugal); the National Priorities Research Program by Qatar National Research Fund; the Ministry of Science and Higher Education, projects no. 0723-2020-0041 and no. FSWW-2020-0008, and the Russian Foundation for Basic Research, project No.19-42-703014 (Russia); MCIN/AEI/10.13039/501100011033, ERDF ``a way of making Europe", and the Programa Estatal de Fomento de la Investigaci{\'o}n Cient{\'i}fica y T{\'e}cnica de Excelencia Mar\'{\i}a de Maeztu, grant MDM-2017-0765 and Programa Severo Ochoa del Principado de Asturias (Spain); the Stavros Niarchos Foundation (Greece); the Rachadapisek Sompot Fund for Postdoctoral Fellowship, Chulalongkorn University and the Chulalongkorn Academic into Its 2nd Century Project Advancement Project (Thailand); the Kavli Foundation; the Nvidia Corporation; the SuperMicro Corporation; the Welch Foundation, contract C-1845; and the Weston Havens Foundation (USA).
\end{acknowledgments}

\bibliography{auto_generated}
\cleardoublepage \appendix\section{The CMS Collaboration \label{app:collab}}\begin{sloppypar}\hyphenpenalty=5000\widowpenalty=500\clubpenalty=5000\cmsinstitute{Yerevan~Physics~Institute, Yerevan, Armenia}
A.~Tumasyan
\cmsinstitute{Institut~f\"{u}r~Hochenergiephysik, Vienna, Austria}
W.~Adam\cmsorcid{0000-0001-9099-4341}, J.W.~Andrejkovic, T.~Bergauer\cmsorcid{0000-0002-5786-0293}, S.~Chatterjee\cmsorcid{0000-0003-2660-0349}, M.~Dragicevic\cmsorcid{0000-0003-1967-6783}, A.~Escalante~Del~Valle\cmsorcid{0000-0002-9702-6359}, R.~Fr\"{u}hwirth\cmsAuthorMark{1}, M.~Jeitler\cmsAuthorMark{1}\cmsorcid{0000-0002-5141-9560}, N.~Krammer, L.~Lechner\cmsorcid{0000-0002-3065-1141}, D.~Liko, I.~Mikulec, P.~Paulitsch, F.M.~Pitters, J.~Schieck\cmsAuthorMark{1}\cmsorcid{0000-0002-1058-8093}, R.~Sch\"{o}fbeck\cmsorcid{0000-0002-2332-8784}, M.~Spanring\cmsorcid{0000-0001-6328-7887}, S.~Templ\cmsorcid{0000-0003-3137-5692}, W.~Waltenberger\cmsorcid{0000-0002-6215-7228}, C.-E.~Wulz\cmsAuthorMark{1}\cmsorcid{0000-0001-9226-5812}
% ...
\vskip\cmsinstskip
% ...
1:~Also at TU Wien, Wien, Austria\\
% ...
\end{sloppypar}
%%% END EDITABLE REGION %%%
% skeleton_end
\end{document}